\documentclass[default]{aastex63}
\usepackage{hyperref}
\usepackage{afterpage}

\newcommand{\lineratio}{\mbox{1--0~S(1)/2--1~S(1)}} 
\newcommand{\hours}{\textsuperscript{h}}
\newcommand{\minutes}{\textsuperscript{m}}
\newcommand{\seconds}{\fs{}}
\newcommand{\degrees}{$^\circ$}
\newcommand{\htwo}{H$_2$} 

\newcommand{\hi}{\ion{H}{1}}
\newcommand{\hii}{\ion{H}{2}}

\begin{document}

\title{A Near-infrared Survey of UV-excited Molecular Hydrogen in Photodissociation Regions}

\author[0000-0001-6909-3856]{Kyle F. Kaplan}
\affiliation{SOFIA Science Center, Universities Space Research Association, NASA Ames Research Center, MS 232-12, Moffett Field, CA 94035, USA}
\affiliation{Department of Astronomy, University of Texas at Austin, 2515 Speedway, Stop C1400, Austin, TX 78712-1205, USA.}
\email{kkaplan@usra.edu}

\author[0000-0002-4017-5572]{Harriet L. Dinerstein}
\affiliation{Department of Astronomy, University of Texas at Austin, 2515 Speedway, Stop C1400, Austin, TX 78712-1205, USA.}
\email{harriet@astro.as.utexas.edu}

\author[0000-0003-4770-688X]{Hwihyun Kim}
\affiliation{Gemini Observatory/NSF's NOIR Lab, Casilla 603, La Serena, Chile}
\email{hkim@gemini.edu}

\author[0000-0003-3577-3540]{Daniel T. Jaffe}
\affiliation{Department of Astronomy, University of Texas at Austin, 2515 Speedway, Stop C1400, Austin, TX 78712-1205, USA.}
\email{dtj@astro.as.utexas.edu}

\accepted{2021 June 02}
\submitjournal{The Astrophysical Journal}

\begin{abstract}
We present a comparative study of the near-infrared (NIR) \htwo{} line
emission from five regions near hot young stars: Sharpless 140, NGC 2023, IC 63, the Horsehead Nebula, and the Orion Bar.  This emission originates in photodissociation or photon-dominated regions (PDRs), interfaces between photoionized and molecular gas near hot (O) stars or reflection nebulae illuminated by somewhat cooler (B) stars.  In these environments, the dominant excitation mechanism for NIR emission lines originating from excited rotational-vibrational (rovibrational) levels of the ground electronic state is radiative or UV excitation (fluorescence), wherein absorption of far-UV photons pumps
\htwo{} molecules into excited electronic states from which they decay into the upper levels of the NIR lines.  Our sources span a range
of UV radiation fields
($G_0 = 10^2$--$10^5$)
and gas densities ($n_H = 10^4$--$10^6$~cm$^{-3}$),
enabling examination of how these properties affect the emergent spectrum.  We obtained high-resolution ($R\! \approx\! 45,\!000$) spectra spanning $1.45$--$2.45$~$\mu$m on the 2.7m Harlan J. Smith Telescope at McDonald Observatory with the Immersion Grating INfrared Spectrometer (IGRINS), detecting up to over 170 transitions per source from excited vibrational states ($v = 1$--$14$).  The
populations of
individual rovibrational levels derived from these data clearly confirm UV excitation.  
Among the five PDRs in our survey, the Orion Bar shows the greatest deviation of the populations and spectrum from pure UV excitation, while Sharpless 140 shows the least deviation.
However, we find that all five PDRs exhibit at least some modification of the level populations relative to their values under pure UV excitation, a result we attribute to collisional effects.
\end{abstract}

\section{Introduction}

Photodissociation or photon-dominated regions (PDRs) arise where non-ionizing far-UV (FUV) photons are the main driver of the physics and chemistry in interstellar gas.
PDRs exist in a wide variety of environments in the interstellar medium (ISM) including the edges of large-scale regions of high-mass star formation and radiation-bounded compact \hii{} regions, reflection nebulae around B stars, and in planetary nebulae where UV light emitted from hot post-AGB stars interacts with previously ejected neutral gas.
In fact, 
most of the diffuse neutral ISM is essentially low-density PDR material \citep{hollenbach91, hollenbach99}.

In regions where massive stars are forming, O and B stars generate plentiful FUV photons that excite and dissociate the surrounding molecular gas. 
When the UV radiation field is steady with time, 
PDRs reach a state of equilibrium with a layered
structure due to the fact that different species
have differing abundances and
absorb UV radiation of different energy ranges \citep{tielens85}.
While extreme-UV (EUV) photons are absorbed by \hi{} gas that marks the edge of the \hii{} region (if present), FUV photons  
easily pass into regions dominated by \hi{} that also contain molecules including \htwo{}  \citep{sternberg88, sternberg14}.  In this region,
absorption of FUV photons in the Lyman and Werner bands electronically excites \htwo{} molecules,
which then either dissociate or decay into excited rotational-vibrational (rovibrational) levels of the ground electronic state \citep{field66, black76}. The dissociated \htwo{} re-forms on dust grains via exothermic catalytic reactions, which eject the \htwo{} back into the gas phase \citep{gould63, hollenbach71}.

\htwo{} lacks a permanent electric dipole moment. 
Molecules in excited rovibrational levels of the ground electronic state can decay to the ground vibrational state only through collisions, or electric quadrupole ($\Delta~J=~0,~\pm2$) transitions that give
rise to a rich spectrum of emission lines with wavelengths ranging from the red end of the optical spectral region to the mid-IR.
Since these lines are optically thin, their fluxes are proportional to the number of molecules in the upper level of each transition. 
Observing many lines enables us to determine the detailed rovibrational level populations,
which reflect the physical conditions within PDRs where the emission arises.

The FUV photons that excite the Lyman and Werner bands of \htwo{} fall within a narrow range of energies from 
$11.2$--$13.6$~eV 
($912 <  \lambda < 1110$~\AA{}).
Consequently, differences in the O or B star
temperature have little effect on the relative \htwo{} level populations \citep{draine96}.
The PDR gas is primarily
heated by collisions with grain photoelectrons and through collisional de-excitation of UV-excited \htwo{}.  
Other mechanisms such as cosmic rays \citep{pellegrini09} and \htwo{} formation \citep{lepetit06, lebourlot12} might also contribute to heating the gas.
The gas cooling is dominated by radiation from atomic and ionic species.
At low densities and temperatures, the emergent spectrum from UV-excited \htwo{} is fairly invariant to detailed properties of the gas and the UV source \citep{black87, sternberg88, draine96}.   
However at higher temperatures and/or densities, collisions between \htwo{} and other particles modify the \htwo{} rovibrational level populations and the emergent spectrum \citep{sternberg89}. 
We can use the effects of collisions on the rovibrational level populations of UV-excited \htwo{} to probe the gas temperature, gas density, and FUV field in a PDR.

A sensitive high-resolution near-IR spectrometer that covers a large wavelength range is the ideal tool for observing \htwo{} rovibrational transitions from many different levels and probing the conditions within PDRs \citep[][hereafter referred to as K17]{le16, kaplan17}.
However, such studies have so far been made for only a few PDRs, notably the reflection nebula NGC 7023 and the Orion Bar.
The true power of this approach can be realized only by applying it to a wider variety of PDRs
that cover a range of UV field intensities, gas densities, and gas temperatures.  
In this paper, we present new observations of four PDRs and compare their \htwo{} spectra
to each other and to the Orion Bar PDR (K17).
This study provides benchmark observations of a set of reference PDRs with a range in these properties, which illustrate and constrain the effects and dependencies of the \htwo{} emission spectrum and offer detailed tests of PDR models.

\section{PDRs in the Survey}

\begin{deluxetable}{c|ccccc}
\tablecaption{PDR Properties and Observations}\label{tab:sf-pdrs}
\tablecolumns{6}
\tablehead{
\colhead{PDR}	&
\colhead{S 140} &
\colhead{NGC 2023} &
\colhead{IC 63} &
\colhead{Horsehead Nebula} &
\colhead{Orion Bar}
}
\startdata
Type &
\hii{} Region Edge	&
Reflection Nebula &
Reflection Nebula &
Dark Nebula Edge &
\hii{} Region Edge \\
Distance (pc)&
$764 \pm 27$\tablenotemark{a} &
$401 \pm 9$\tablenotemark{b} &
$188 \pm 28$\tablenotemark{c} &
$344 \pm 25$\tablenotemark{d} &
$450 \pm 50$\tablenotemark{e} \\
Illuminating Star &
HD 211880 &
HD 37903 &
$\gamma$ Cas &
$\sigma$ Ori &
$\theta^1$~Ori~C \\
Spectral Type &
B0.5V\tablenotemark{f}  &
B1.5V\tablenotemark{g} &
B0.5IVpe\tablenotemark{h} &
O9.5V\tablenotemark{i}  &
O7V\tablenotemark{j}  \\
$T_{\star,\mbox{eff}}$ (K) &
29,000\tablenotemark{f} &
23,700\tablenotemark{g} &
25,000\tablenotemark{k} &
33,000\tablenotemark{l} &
39,000\tablenotemark{j} \\
$G_0$ &
$4.0\times10^{2}$\tablenotemark{f} &
$1.7 \times 10^{4}$\tablenotemark{m} &
$1.5 \times 10^{2}$\tablenotemark{n} &
$10^{2}$\tablenotemark{o} &
$3\times10^{4}$\tablenotemark{p} \\
$n_H$ (cm $^{-3}$) &
$10^{4}$\tablenotemark{f} &
$2 \times 10^{5}$\tablenotemark{m} &
$1.3 \times 10^{4}$\tablenotemark{n}&
$2 \times 10^{4}$\tablenotemark{o} &
$10^{6}$\tablenotemark{q} \\
$G_0/n_H$ (cm$^{3}$) &
$4.0\times10^{-2}$ &
$8.5\times10^{-2}$ &
$1.2\times10^{-2}$ &
$5.0\times10^{-3}$ &
$3.0\times10^{-2}$ \\
\hline
R.A. (J2000) &
22\hours{}19\minutes{}13\seconds{}40 &
05\hours{}41\minutes{}37\seconds{}71 &
00\hours{}59\minutes{}02\seconds{}74 &
05\hours{}40\minutes{}53\seconds{}60 &
05\hours{}35\minutes{}19\seconds{}73 \\
Decl. (J2000) &
$+$63\degrees{}17\arcmin{}53\arcsec{}9	&
$-$02\degrees{}16\arcmin{}50\arcsec{}0 	&
$+$60\degrees{}53\arcmin{}06\arcsec{}7	&
$-$02\degrees{}28\arcmin{}00\arcsec{}0	&
$-$05\degrees{}25\arcmin{}26\arcsec{}7 \\
Date Obs. (UT) &
2015 Nov 02 &
2015 Nov 03 &
2015 Nov 03 &
2015 Nov 01&
2014 Oct 24 \\
Exp. Time (min.) &
100 &
70 &
80 &
55 &
30 \\
P.A. (\degrees) &
45 &
15 &
45 &
76 &
135 \\
Std. Star &
HR 8422 &
HR 1724 &
HR 598 &
HR 1482 &
HD 34317 \\
\hline
$A_K$\tablenotemark{1} &
0.52 &
0.45 &
0.30 &
0.70 &
0.72 \\
O/P\tablenotemark{2} &
$1.41 \pm 0.12$ &
$2.17 \pm 0.24$ &
$1.83 \pm 0.21$ &
$1.34 \pm 0.16$ &
$2.99 \pm 0.38$ \\
1--0~S(1)/2--1~S(1)\tablenotemark{3} &
$2.41 \pm 0.32$ &
$3.10 \pm 0.44$ &
$2.92 \pm 0.48$ &
$2.86 \pm 0.33$ &
$5.40 \pm 0.61$  \\
\enddata
\tablerefs{
$^a$\citet{hirota08},
$^b$\citet{gaia18},
$^c$\citet{perryman97},
$^d$\citet{caballero08},
$^e$\citet{schlafly14},
$^f$\citet{timmermann96},
$^g$\citet{compiegne08},
$^h$\citet{shenavrin11},
$^i$\citet{warren77},
$^j$\citet{ferland12},
$^k$\citet{sigut07},
$^l$\citet{panagia73},
$^m$\citet{sheffer11},
$^n$\citet{andrews18},
$^o$\citet{habart05},
$^p$\citet{marconi98},
$^q$\citet{goicoechea16}
}
\tablenotetext{1}{
The best-fit K-band dust extinction in magnitudes ($A_K$),
used to extinction correct \htwo{} line fluxes for each PDR, as described in Section \ref{sect:datareduction}.
}
\tablenotetext{2}{
The best-fit ortho-to-para ratio (O/P) observed for the rovibrationally excited \htwo{}, as described in Sections \ref{sect:measure-ortho-to-para} \& \ref{sect:ortho-to-para-ratio-results}.
}
\tablenotetext{3}{
Flux ratio of the \htwo{} 1--0~S(1) to 2--1~S(1) lines in each PDR, as described in $\S$ \ref{sec:sfr-pdrs-h2-line-ratio}.  The individual line fluxes are reported in Table \ref{tab:sf-pdr-flux}.
}
\end{deluxetable}

\begin{figure}
\includegraphics[width=0.5\textwidth]{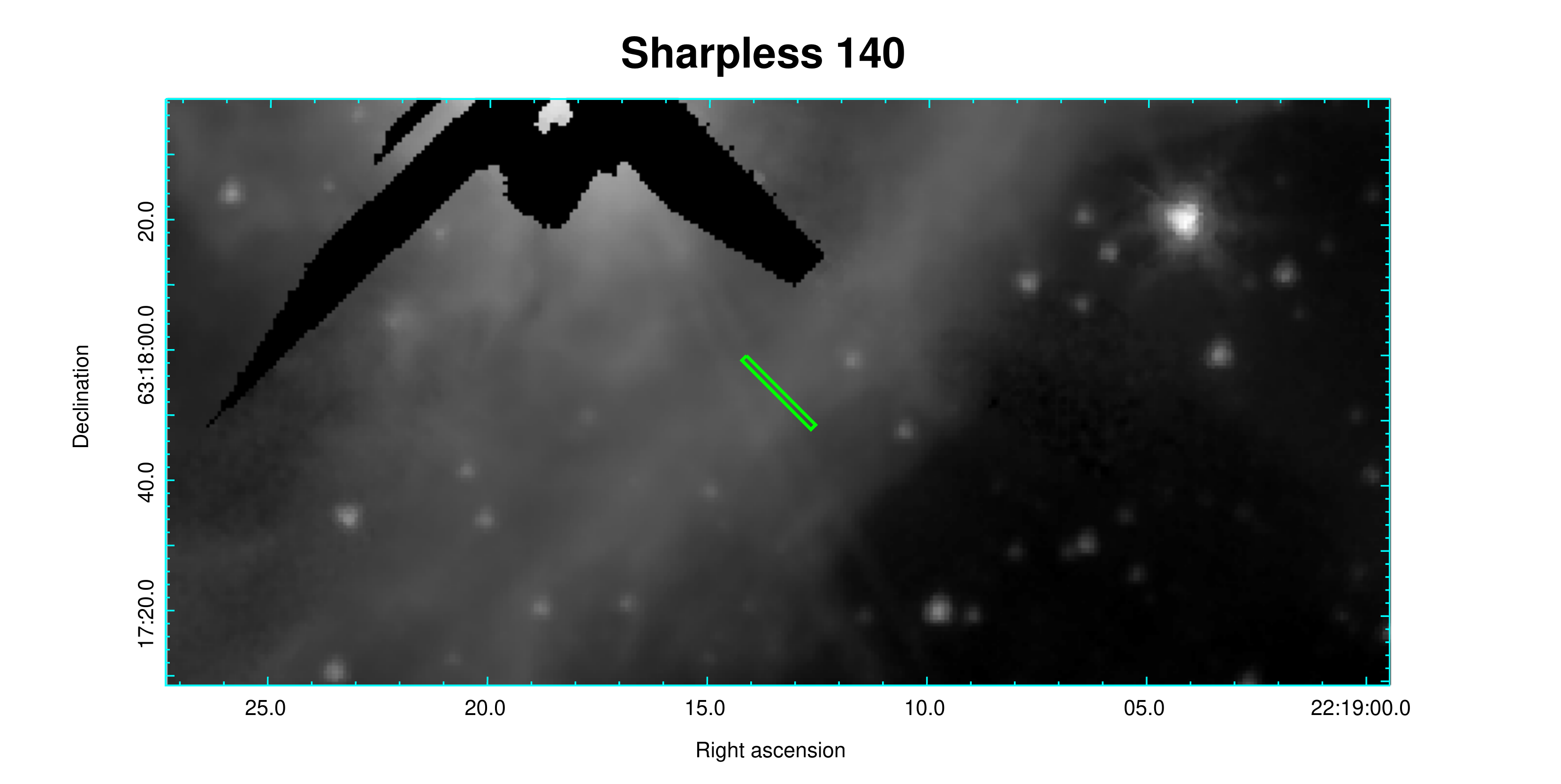}
\includegraphics[width=0.5\textwidth]{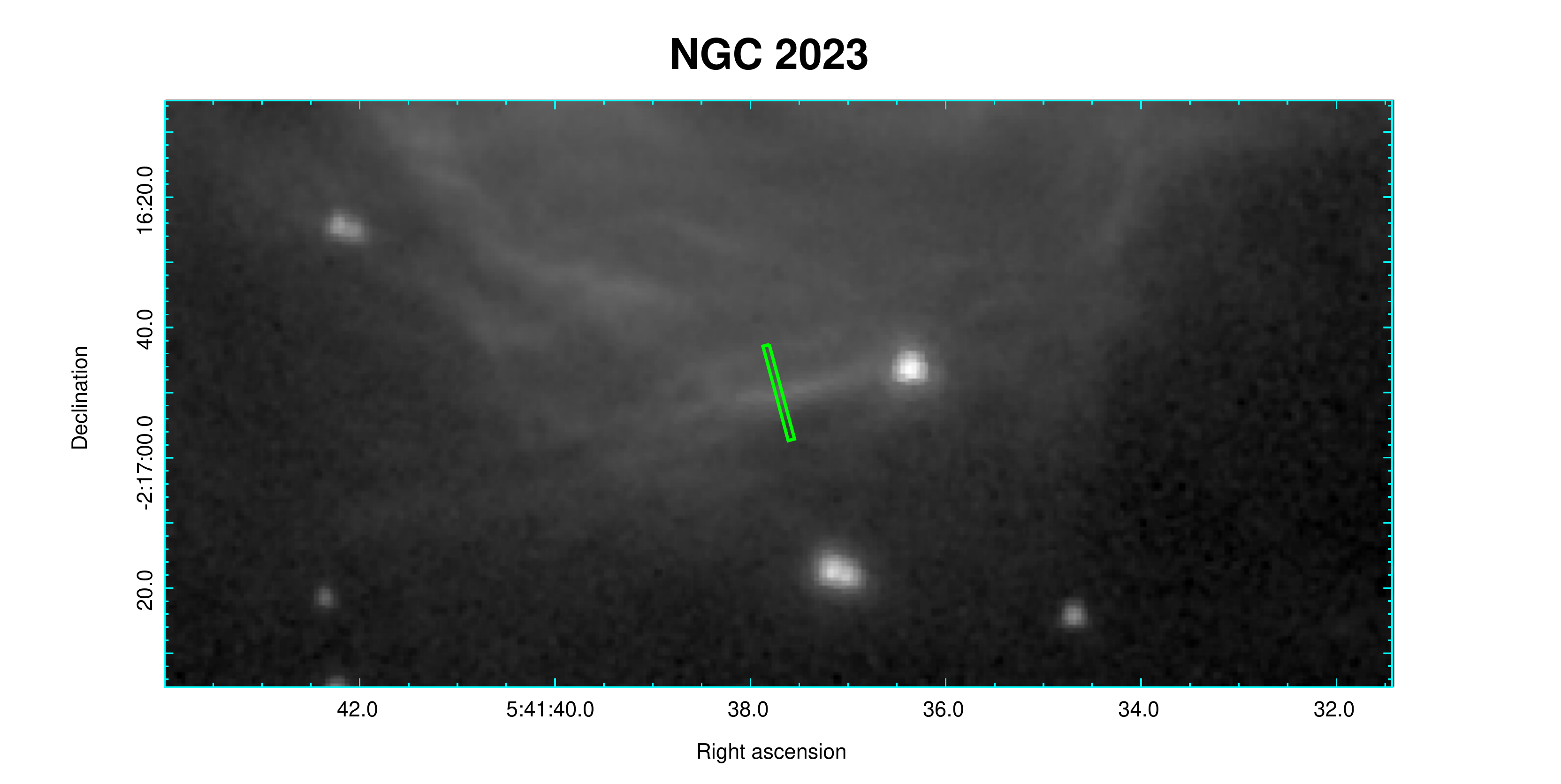}
\includegraphics[width=0.5\textwidth]{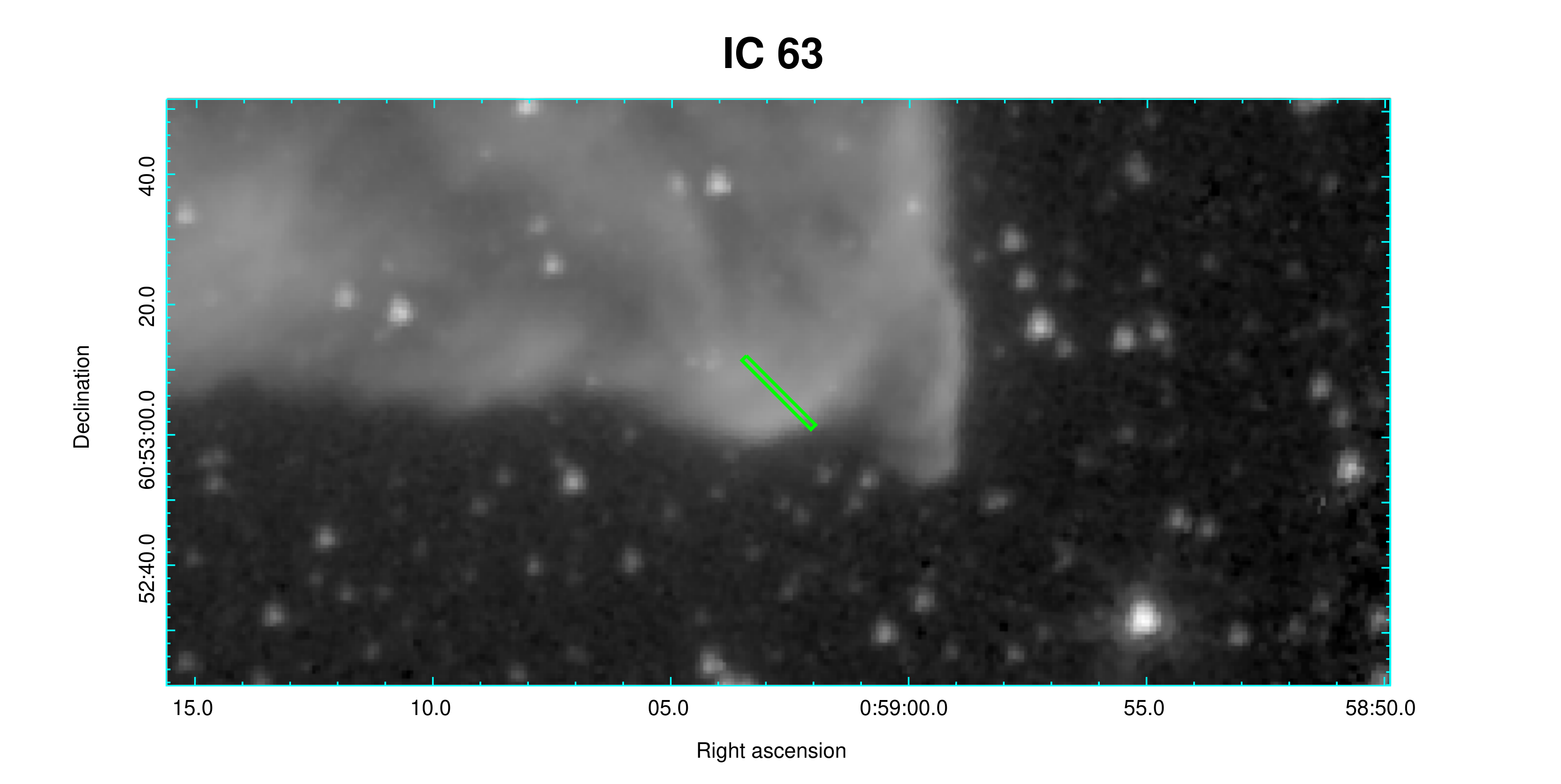}
\includegraphics[width=0.5\textwidth]{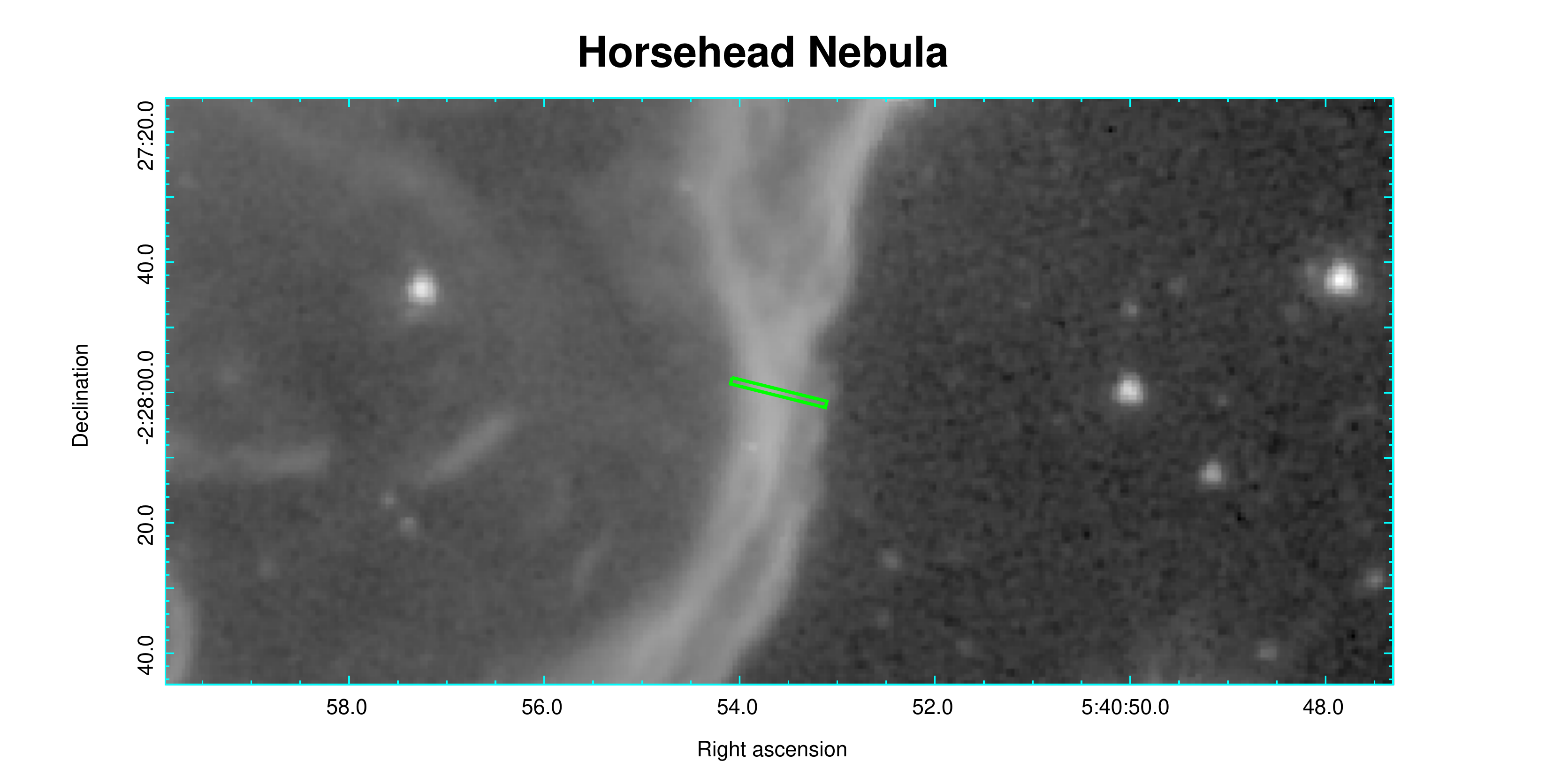}
\includegraphics[width=0.5\textwidth]{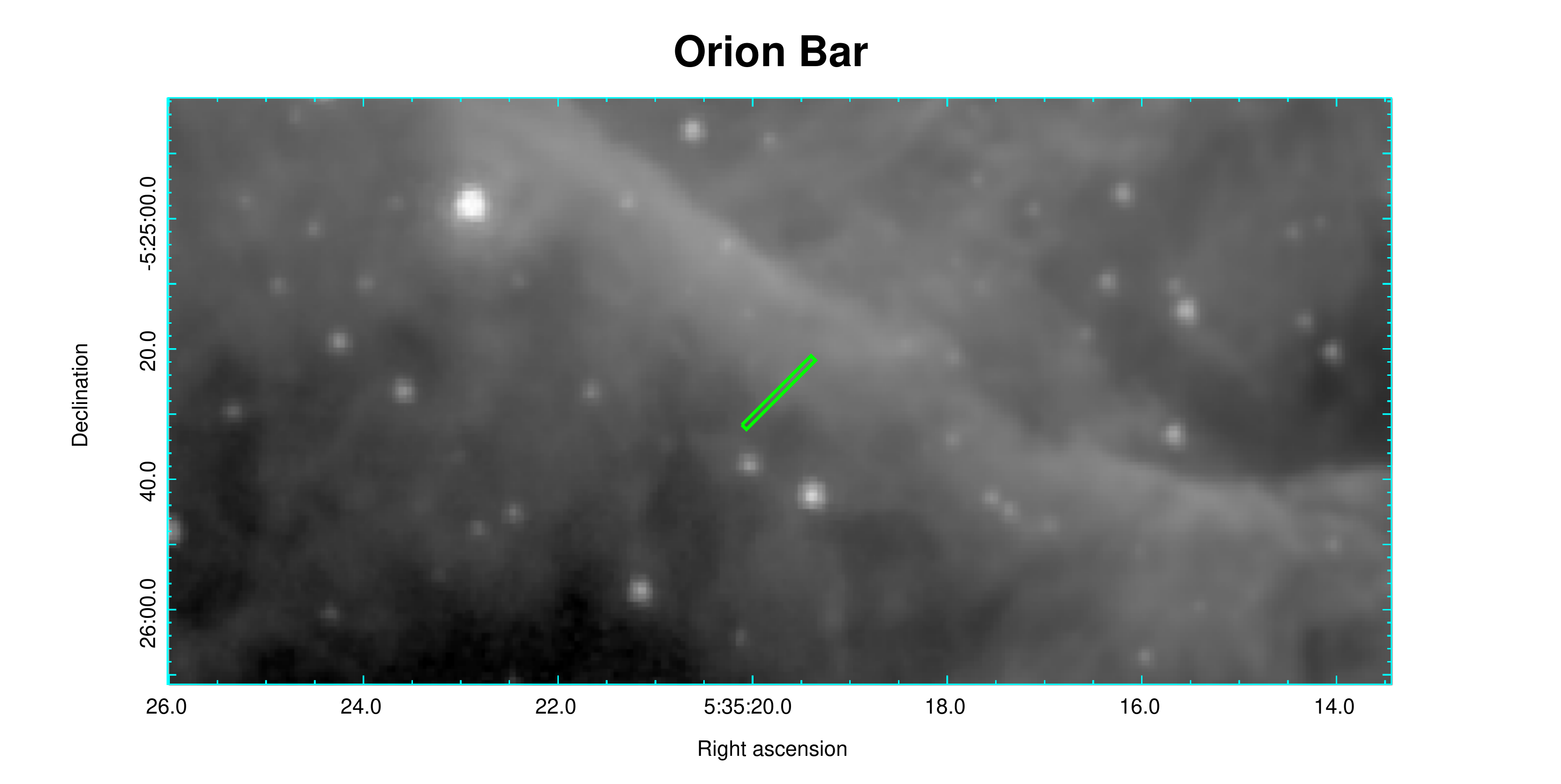}
\caption{
Finder charts showing the locations of the IGRINS slit for each of the observed PDRs in the survey.   The green rectangle represents the size and position angle of the 1\arcsec{}$\times$15\arcsec{} IGRINS slit.  North is up and east is  toward the left.  
The images used in these finder charts are Spitzer IRAC 3.6~$\mu$m \citep{fazio04} super-mosaics generated as Spitzer Enhanced Image Products from archival data and were obtained from the IRSA Spitzer Heritage Archive.  The black portions of the Sharpless 140 finder chart are due to overexposure.
}
\label{fig:finder-charts}
\end{figure}

We selected Sharpless~140, NGC~2023, IC~63, and the Horsehead Nebula PDRs for this survey based on their observability in the northern sky, previous detections of UV-excited \htwo{}, and to ensure a set of targets with a range of properties (UV field, density, temperature, etc.).  We also include our data and results for the Orion Bar from K17.  Table~\ref{tab:sf-pdrs} summarizes the observations and physical properties of the PDRs in this survey.
We report the physical parameters from the literature $G_0$ which is a dimensionless measure of the UV intensity normalized to the interstellar radiation field of \citet{habing1968} and $n_H$ which is the gas density.  We also report the ratio of the UV intensity to the gas density $G_0/n_H$ which parameterizes the effects of UV photons upon the gas.
  
All targets were observed with the Immersion GRating INfrared Spectrometer (IGRINS) on the 2.7 m Harlan J. Smith telescope at McDonald Observatory.  IGRINS is a high-resolution ($R\! \sim\! 45,\! 000$) cross-dispersed echelle spectrometer with separate near-IR channels for the $H$ and $K$ bands that provides simultaneous observations of a full wavelength coverage from $1.45$--$2.45$ $\mu$m  \citep{park14}.  The IGRINS slit subtends  $1\arcsec \times 15\arcsec$ on the sky and can be rotated to any position angle (P.A.; measured counterclockwise from NE).
All PDRs were observed by alternating the slit between the target position and a blank region of the sky approximately one degree offset from the target, with equal amounts of time spent on target and sky.  An A0V standard star was observed at a similar airmass to each target for telluric absorption line correction and relative spectrophotometric flux calibration.
Table~\ref{tab:sf-pdrs} lists the coordinates of the IGRINS slit positions, date of the observations, P.A.s, integration times, and A0V standard stars used for all our observations.  
Figure \ref{fig:finder-charts} shows finder charts and the IGRINS slit positions for all targets.
Below we describe the observing details for each PDR included in the survey.

\subsection{S 140}

Sharpless 140 (S~140) is an \hii{} region surrounding a cluster of several B stars that includes some reflection nebulosity in the background.  The brightest UV source is the B0.5V star HD~211880 \citep{timmermann96}.  The geometry of the S~140 PDR is similar to that of the Orion Bar (i.e. it is viewed nearly edge-on),
but with a less intense FUV radiation field.  \htwo{} emission was observed from S~140 with the Infrared Space Observatory by \citet{timmermann96} and \citet{habart04}.
We placed the IGRINS slit
on the edge of the bright near-IR filament in S~140 with the slit rotated to a PA of 45\degrees{} (measured counter-clockwise from NE), perpendicular to the \hi{}/\htwo{} dissociation interface.

\subsection{NGC 2023}

NGC~2023 is a reflection nebula surrounding the B1.5V star HD~37903 where UV radiation has carved out a bubble within the L1630 molecular cloud \citep{compiegne08}.  It is one of the brightest and best-studied PDRs.  There have been many studies of its UV-excited \htwo{} emission: \citet{sellegren86}, \citet{hasegawa87}, \citet{takayanagi87}, \citet{gatley87b},
\citet{black87},
\citet{sternberg88}, \citet{sterberg89b}, \citet{tanaka89}, \citet{hippelein89}, \citet{howe90}, \citet{burton90b}, \citet{burton92b}, \citet{field94}, \citet{field98}, \citet{burton98}, \citet{martini99}, \citet{takami00}, \citet{habart04}, \citet{fleming10}, \citet{habart11}, and \citet{sheffer11}.  We placed the IGRINS slit on the bright southern ridge
with a PA of 135\degrees{} perpendicular to the edge of the ridge.

\subsection{IC 63}

IC~63 is a nearby ($188 \pm 28$ pc) reflection nebula near the B0.5IVpe star $\gamma$~Cas \citep{perryman97, shenavrin11}.  The \htwo{} emission in this source has been the subject of numerous studies including \citet{witt89}, \citet{sternberg89}, \citet{luhman97}, \citet{hurwitz98}, \citet{habart04}, \citet{france05}, \citet{thi09}, and \citet{fleming10}.
In $K$-band images, the PDR forms a ``V''-shaped cloud with the tip of the V pointing southwest toward $\gamma$~Cas.  We positioned the IGRINS slit
on the brightest part of the tip of the nebula nearest to $\gamma$~Cas with the slit at a P.A. of 45\degrees{}, perpendicular to the \hi{}/\htwo{} dissociation interface.

\subsection{Horsehead Nebula}

The Horsehead Nebula is a dark portion of the L1630 molecular cloud.  The PDR arises where UV light from the O9.5V star $\sigma$ Ori \citep{warren77} falls  upon the western-facing edge of the ``head.''  
The UV field intensity for the Horsehead PDR is the lowest of our sample ($G_0 \sim 100$; \citealt{abergel2003}), since the illuminating star is relatively distant ($\approx 3.5$ pc; \citealt{abergel2003}).
Like the Orion Bar and S 140, the Horsehead PDR is viewed nearly edge-on.  
The \htwo{} 1--0~S(1) line emission was mapped by \citet{habart05} and the pure rotational lines were mapped by \citet{habart11}.
We placed the IGRINS slit at the location of the brightest \htwo{} 1--0~S(1) line emission,
with P.A.~=~76\degrees{}, which is perpendicular to the \hi{}/\htwo{} dissociation interface.

\subsection{Orion Bar}
The Orion Bar is the southeastern illuminated edge of the Orion Nebula  \hii{} region surrounding the Trapezium star cluster in which the strongest UV source is the O7V star $\theta^1$~Ori~C. e 
The observations of the Orion Bar included here are from K17, which covers our \htwo{} observations and the literature in great detail.
 The Orion Bar has a density of $n_H \sim10^6$~cm$^{-3}$ \citep{goicoechea16} and a UV intensity of $G_0 \sim 3.0\times10^4$ \citep{marconi98}, which is the highest density and UV intensity out of all the PDRs in this survey.   
 We positioned the IGRINS slit on the point of brightest \htwo{} emission with a P.A. of 135\degrees{}, perpendicular to the \hi{}/\htwo{} dissociation interface.

\section{Analysis}

\subsection{Data Reduction, Extinction Correction, and Line Flux Measurements} \label{sect:datareduction}

Our data reduction, calibration, and line flux extraction procedures are described in K17.
We use the public IGRINS Pipeline Package  \citep{plp}\footnote{IGRINS Pipeline Package (PLP): \url{https://github.com/igrins/plp}} for data reduction, after removing cosmic rays from the raw frames using the Python implementation of LA--Cosmic \citep{dokkum2001}\footnote{Python implementation of LA--Cosmic by Malte Tewes: \url{https://obswww.unige.ch/~tewes/cosmics_dot_py/}}.  
We use the code ``plotspec,''\footnote{``Plotspec:'', \url{https://github.com/kfkaplan/plotspec}}, which is designed for processing 2D IGRINS spectra of extended emission, for our data analysis. 
The 2D orders from the echellograms are combined into one long 2D spectrum.
A0V standard stars are used for simultaneous telluric absorption line correction and relative spectrophotometric flux calibration. 
Lines are identified using the list from \citet{roueff19}.
The 2D line profiles are linearly interpolated into position-velocity space, and line fluxes are extracted using a weighted optimal extraction routine with the profile of the 1--0~S(1) line used to establish the weights.
We add a 15\% systematic uncertainty to all of the extracted line fluxes to account for uncertainty in the relative flux calibration, telluric correction, and extinction correction. 

The extracted line fluxes for each PDR are extinction corrected using the best-fit $A_K$, the dust extinction in the $K$-band.
The NIR extinction law can be approximated as a power law of the form $A_{\lambda} \propto \lambda^{-\beta}$.  Historical values of $\beta$ range between $1.6$--$1.8$ \citep[see][and references therein]{draine03}.  More recent studies by \citet{nishiyama06, nishiyama09} and \citet{wang19} suggest a higher value of $\beta \sim 2$.  To correct for dust extinction, we have selected $\beta = 1.8$.
Setting  $\beta=1.6$ or $\beta=2.0$ gives $A_K$ values that are $\sim\! 15 \%$ larger or smaller, respectively.
While the value of $\beta$ is somewhat uncertain, $A_K$ values for these sources are sufficiently small that the differential extinction corrections are not a significant source of uncertainty.  Changing the value  for $\beta$ has a $<\! 1\%$ effect on our measured line flux ratios. 
For each PDR, we find the value of $A_K$ that best matches the intrinsic flux ratios for every pair of \htwo{} emission lines that arise from the same upper $v$ and $J$ state to their observed flux ratios and report the best-fit $A_K$ in Table~\ref{tab:sf-pdrs}.
For example, 3--1~O(5) at 1.522033~$\mu$m and 3--2~S(1) at 2.386547~$\mu$m both arise from the $v=3$, $J=3$ state.  There are between 13--66 such pairs found in the spectra for our PDR sample that we can use as part of our determination of $A_K$.

\begin{figure}
\includegraphics[width=1.0\textwidth]{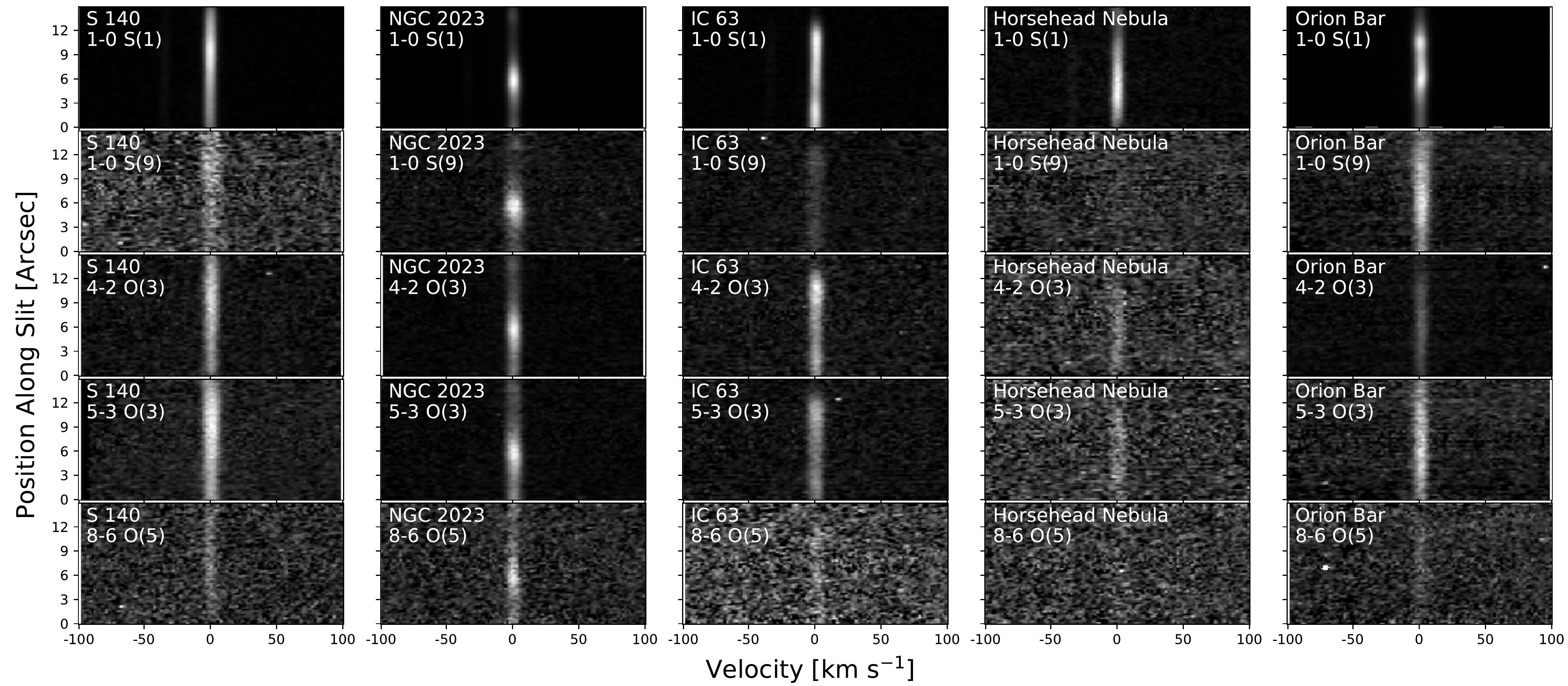}
\caption{Position-velocity diagram thumbnails of a representative set of \htwo{} emission lines seen in all the PDRs in our survey.  These transitions arise from upper states with a wide range of excitation energies.  The central slit position for each PDR is reported in Table~\ref{tab:sf-pdrs}.  The velocity zero-point for each PDR is centered on the 1--0~S(1) line.}
\label{fig:sf-pdr-2d}
\end{figure}

\begin{figure}
\includegraphics[width=1.0\textwidth]{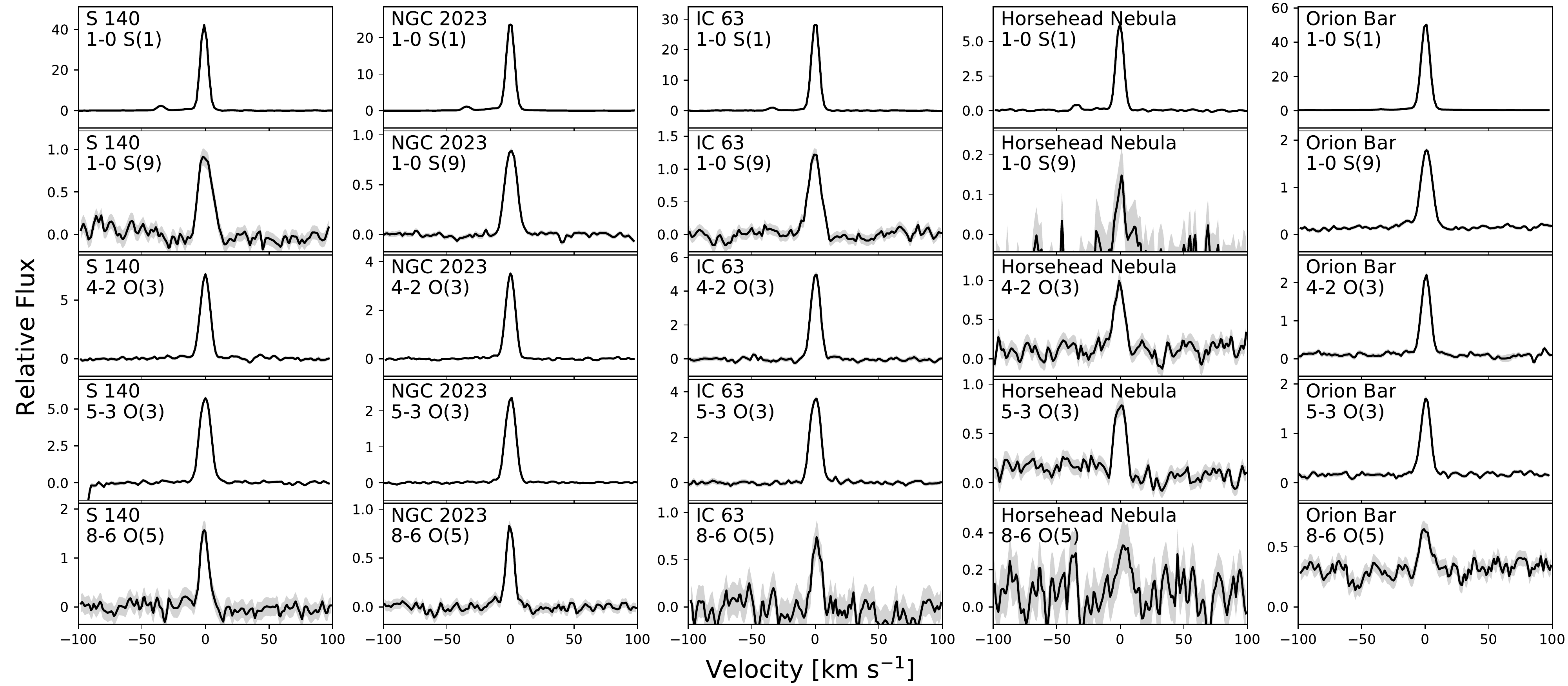}
\caption{1D thumbnails of the \htwo{} lines in Figure~\ref{fig:sf-pdr-2d}.  The 1D spectra are created by summing all of the pixels along the slit in position space.  The gray shaded region represents the $\pm1\sigma$ statistical uncertainty.  Fluxes are scaled to facilitate comparison between different lines in a given PDR.}
\label{fig:sf-pdr-1d}
\end{figure}

We detect a large number of lines in each PDR.  
Figures \ref{fig:sf-pdr-2d} and \ref{fig:sf-pdr-1d} show 2D and 1D thumbnails from a representative set of \htwo{} lines.
The lines are all quite narrow, corresponding to velocity widths of only a few km kilometers per second.
This is consistent with the \htwo{} being relatively cool and UV-excited, as opposed to arising from shocks.
In shocks, we would expect to see larger velocity widths from gas motion (unless there is unfavorable alignment) and thermal broadening.  The observed velocity widths place an upper limit on the gas temperature of $\le\! 500$~K, lower than the expected value for shocked gas, which is typically a few thousand kelvin.
Table~\ref{tab:sf-pdr-flux} lists the extracted line fluxes for each PDR normalized to the line flux of 4--2~O(3), which is a fairly bright line seen in all of the PDRs and is uncontaminated by telluric absorption or OH emission lines.
The fluxes are reported as ratios to this reference line due to variations in airmass, seeing, and observing conditions that resulted in uncertainties that prevented us from determining absolute fluxes.

\begin{deluxetable}{ccccccccc}
\centerwidetable
\tablecolumns{9}
\tabletypesize{\footnotesize}
\tablecaption{\htwo{} Line Fluxes for Our PDRs (Normalized to the 4--2 O(3) Line)}\label{tab:sf-pdr-flux}
\tablehead{\colhead{Line ID} & \colhead{$\lambda$ ($\mu$m)} & \colhead{$v_u$} & \colhead{$J_u$} & \colhead{S 140} & \colhead{NGC 2023} & \colhead{IC 63} & \colhead{Horsehead Nebula} & \colhead{Orion Bar}}
\startdata
1--0 Q(1) & 2.406592 & 1 & 1 & $ 4.357 \pm  0.446$ & $ 7.910 \pm  0.853$ & $ 5.229 \pm  0.630$ & $ 5.288 \pm  0.475$ & $23.334 \pm  2.065$ \\ 
1--0 Q(2) & 2.413439 & 1 & 2 & $ 2.412 \pm  0.248$ & $ 2.640 \pm  0.285$ & $ 2.276 \pm  0.275$ & $ 3.131 \pm  0.283$ & $ 5.539 \pm  0.492$ \\ 
1--0 S(0) & 2.223290 & 1 & 2 & $ 2.169 \pm  0.210$ & $ 2.371 \pm  0.243$ & $ 2.111 \pm  0.246$ & $ 3.061 \pm  0.254$ & $ 5.208 \pm  0.425$ \\ 
1--0 S(1) & 2.121834 & 1 & 3 & $ 3.740 \pm  0.348$ & $ 5.402 \pm  0.536$ & $ 4.336 \pm  0.494$ & $ 4.082 \pm  0.322$ & $15.214 \pm  1.177$ \\ 
1--0 Q(3) & 2.423730 & 1 & 3 & $ 2.387 \pm  0.246$ & $ 3.414 \pm  0.370$ & $ 2.825 \pm  0.341$ & $ 2.890 \pm  0.262$ & $10.160 \pm  0.905$ \\ 
1--0 Q(4) & 2.437489 & 1 & 4 & $ 1.114 \pm  0.115$ & $ 1.211 \pm  0.132$ & $ 1.092 \pm  0.132$ & $ 1.371 \pm  0.126$ & $ 2.506 \pm  0.225$ \\ 
1--0 S(2) & 2.033758 & 1 & 4 & $ 1.922 \pm  0.172$ & $ 2.135 \pm  0.205$ &  --  & $ 2.363 \pm  0.177$ & $ 4.225 \pm  0.310$ \\ 
1--0 S(3) & 1.957559 & 1 & 5 & $ 2.438 \pm  0.210$ & $ 4.124 \pm  0.383$ & $ 3.285 \pm  0.358$ & $ 2.632 \pm  0.189$ & $ 8.835 \pm  0.617$ \\ 
1--0 Q(6) & 2.475559 & 1 & 6 & $ 0.396 \pm  0.042$ & $ 0.466 \pm  0.052$ & $ 0.475 \pm  0.060$ & $ 0.573 \pm  0.074$ &  -- \\
1--0 S(6) & 1.788050 & 1 & 8 &  --  &  --  &  --  &  --  & $ 0.869 \pm  0.053$ \\ 
1--0 S(7) & 1.747955 & 1 & 9 & $ 0.367 \pm  0.028$ & $ 0.779 \pm  0.065$ & $ 0.634 \pm  0.065$ & $ 0.310 \pm  0.028$ & $ 2.012 \pm  0.118$ \\ 
1--0 S(8) & 1.714738 & 1 & 10 & $ 0.094 \pm  0.009$ & $ 0.174 \pm  0.015$ & $ 0.152 \pm  0.018$ &  --  & $ 0.393 \pm  0.023$ \\ 
1--0 S(9) & 1.687761 & 1 & 11 & $ 0.122 \pm  0.009$ & $ 0.267 \pm  0.021$ & $ 0.229 \pm  0.023$ & $ 0.167 \pm  0.014$ & $ 0.649 \pm  0.036$ \\ 
1--0 S(10) & 1.666475 & 1 & 12 & $ 0.027 \pm  0.003$ & $ 0.057 \pm  0.005$ & $ 0.043 \pm  0.006$ &  --  & $ 0.094 \pm  0.006$ \\ 
1--0 S(11) & 1.650413 & 1 & 13 &  --  & $ 0.046 \pm  0.004$ & $ 0.050 \pm  0.006$ &  --  & $ 0.083 \pm  0.005$ \\ 
\enddata
\tablecomments{Undetected lines are marked with a dash.\\
(This table is available in its entirety in machine-readable form in the arXiv source.)}
\end{deluxetable}


\begin{deluxetable}{cccccc}
\tablecaption{Physical Constants for \htwo{} Lines} \label{tab:h2_physical_constants}
\tablecolumns{6}
\tablehead{\colhead{\htwo{} Line ID} & \colhead{$\lambda$} & \colhead{$v_u$} & \colhead{$J_u$} & \colhead{$E_u/k$} & \colhead{$\log_{10}\left(A_{ul}\right)$}}
\startdata
1--0 Q(1) & 2.406592 & 1 & 1 &  6149 & -6.37 \\
1--0 Q(2) & 2.413439 & 1 & 2 &  6471 & -6.52 \\
1--0 S(0) & 2.223290 & 1 & 2 &  6471 & -6.60 \\
1--0 S(1) & 2.121834 & 1 & 3 &  6951 & -6.46 \\
1--0 Q(3) & 2.423730 & 1 & 3 &  6951 & -6.55 \\
1--0 Q(4) & 2.437489 & 1 & 4 &  7584 & -6.57 \\
1--0 S(2) & 2.033758 & 1 & 4 &  7584 & -6.40 \\
1--0 S(3) & 1.957559 & 1 & 5 &  8365 & -6.38 \\
1--0 Q(6) & 2.475559 & 1 & 6 &  9286 & -6.59 \\
1--0 S(6) & 1.788050 & 1 & 8 & 11521 & -6.45 \\
1--0 S(7) & 1.747955 & 1 & 9 & 12817 & -6.53 \\
1--0 S(8) & 1.714738 & 1 & 10 & 14220 & -6.63 \\
1--0 S(9) & 1.687761 & 1 & 11 & 15721 & -6.77 \\
1--0 S(10) & 1.666475 & 1 & 12 & 17311 & -6.98 \\
1--0 S(11) & 1.650413 & 1 & 13 & 18979 & -7.27 \\
\enddata
\tablecomments{Values for $\lambda$, $E_u$, and $A_{ul}$ are from \citet{roueff19}.\\
(This table is available in its entirety in machine-readable form in the arXiv source.)}
\end{deluxetable}


\subsection{\htwo{} Level Populations, Excitation Diagrams, and the Effects of Collisions}\label{sec:h2levelpops}

We calculate the \htwo{} rovibrational level populations from the column densities of the upper state $N_{u}$ of each observed rovibrational transition from
\begin{equation} \label{eq:col-dens}
N_u = {F_{ul} \over \Delta E_{ul} h c A_{ul}},
\end{equation}
where $F_{ul}$ is the flux of the transition from upper ($u$) to lower ($l$) rovibrational states, $\Delta E_{ul}$ is the difference in energy between the states in units of cm$^{-1}$, $A_{ul}$ is the transition probability in s$^{-1}$, $h$ is Planck's constant, and $c$ is the speed of light.  
Table~\ref{tab:h2_physical_constants} lists the physical constants that we used to calculate the level populations.  We use the latest theoretical molecular data on \htwo{} from \citet{roueff19} for the values of $\lambda$, $E_u$, and $A_{ul}$, 
which improves upon the values we previously used in K17.
We normalize all of the column densities to the population of the $v=4$, $J=1$ level derived from the 4--2~O(3) transition,
the same transition we used to normalize the line fluxes.
Table~\ref{tab:sf-pdr-coldens} gives the derived column densities of \htwo{} in the upper levels of each transition, $N_u$, relative to that of the reference level $N_r \equiv N(v=4, J=1)$ in each PDR.  It also lists the ratios of the normalized column densities to those predicted for a ``pure'' UV-excited Cloudy model, in which it is assumed there is negligible collisional modification of the level populations, denoted by $N_m$ (see \S~\ref{sec:compare-to-model} for a detailed discussion of this model).

\begin{figure}
\includegraphics[width=\textwidth]{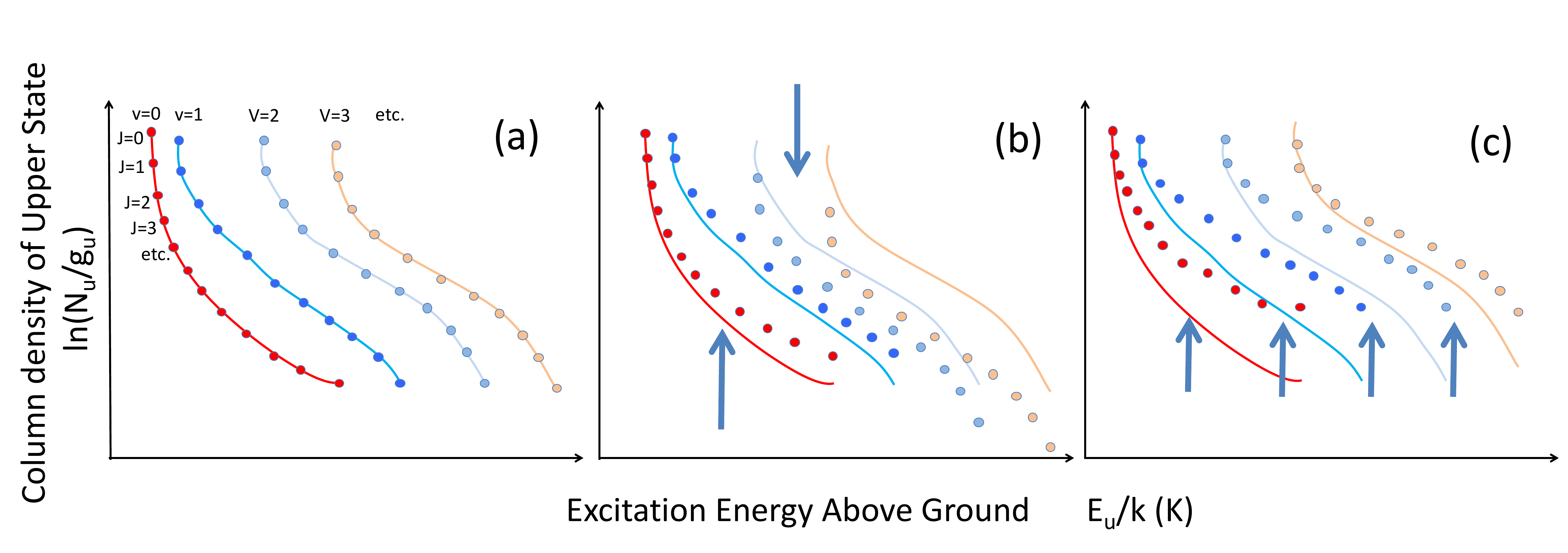}
\caption{Schematic excitation diagrams of
the \htwo{} rovibrational level populations for UV-excited \htwo{}.  Panel (a) depicts the characteristic
nonthermal distribution of unmodified UV excitation.  The solid colored lines represent this pure UV-excited case across all panels.
The other two panels illustrate the effects of increasing thermalization
where collisions modify these level populations from the unmodified UV-excited case.
Panel (b) depicts how collisional excitation raises the relative populations of low excitation energy levels and collisional de-excitation suppresses the populations of high energy levels.
The arrows show the direction the level populations at low and high excitation energy levels move relative to the unmodified UV-excited case.
Panel (c) shows how the relative populations increase at high $J$ across all $v$ levels with increasing thermalization.  
The arrows show how the populations at higher $J$ rise relative to the unmodified UV-excited case.
See Section \ref{sec:h2levelpops} for a detailed discussion.
}
\label{fig:collisions}
\end{figure}

Competition among UV radiative excitation, spontaneous decays, and 
collisional excitation and de-excitation gives rise to two limiting cases for the \htwo{} rovibrational level populations: the thermal limit and the case of ``pure" UV radiative  excitation (fluorescence).  
Figure~\ref{fig:collisions} illustrates a convenient diagnostic tool for determining excitation mechanisms, the excitation diagram, which plots the logarithmic level populations normalized by the statistical weight, $\ln\left(N_u/g_u\right)$, against the excitation energy $E_u/k$ of the upper level of each transition.
In dense 
gas, such as in shocks, the rovibrational levels are excited and de-excited by frequent collisions, and the level populations approach a thermal distribution.
In that limit, the plotted level populations follow a monotonic decline with a (constant)
linear slope of $-(1/T)$ where $T$ is the kinetic temperature of the gas.
On the other hand, in
low-density
gas exposed to UV radiation, the level populations take on a distinctive nonthermal distribution with much larger populations in the $v \ge 1$ levels.
Instead of a single trend that 
declines monotonically with increasing excitation energy, the populations on the excitation diagram are split into different rotation ladders (levels of constant $v$) with declining trends that are roughly parallel with each other.
Observations of the rovibrational line flux ratios readily distinguish between these two limiting cases.
However, in practice, many sources show excitation diagrams intermediate between the UV-excited and thermal limits. 
This could result either from collisional modification of the level populations following UV excitations (e.g., \citealt{sternberg89, burton90}), or from the superposition of spatially distinct UV-excited and thermal components (e.g., \citealt{davis03}).  Distinguishing between these possibilities may be difficult with limited data, but this may be remedied with the use of detailed high-spectral resolution observations.

Since the collision energies depend on the gas temperature and collision rates depend on both the gas density and temperature, subtle modifications of UV-excited \htwo{} rovibrational level populations by collisions can be exploited to probe the physical conditions within a PDR.
Panel (a) in Figure~\ref{fig:collisions} depicts the
nonthermal distribution
of UV-excited \htwo{} on the excitation diagram.
Each excited rovibrational level has a (somewhat temperature dependent) critical density above which the rate of collisional de-excitation is greater than the rate of radiative de-excitation.
At gas densities higher than the critical density \citep[typically $\sim 10^4$--$10^5$~cm$^{-3}$,][]{sternberg89}, collisions will dominate the excitation and de-excitation of a level, driving the level population toward the thermal limit.

As collision rates increase due to the gas getting warmer or denser, all rovibrational levels become increasingly thermalized.
Levels with lower energies ($v=0$~and~$1$)
will experience increasing rates of both collisional excitation and de-excitation,
while higher-$v$ levels are more strongly affected by collisional de-excitations than collisional excitations due to their higher threshold energies, decreasing the population of higher-energy levels relative to lower-energy levels.
UV excitation always pumps more molecules into higher-$v$ levels than the process of thermalization, because \htwo{} can never realistically reach kinetic temperatures high enough to significantly populate the high-$v$ levels without dissociating.  As collision rates, thus thermalization, increase, the number of molecules at high $v$ will decrease.  The end result is that the ``vibrational temperature,'' $T_{\rm vib}$, the excitation temperature determined from the populations across different vibrational levels in the same rotation state, decreases.
Panel (b) in Figure~\ref{fig:collisions} illustrates this effect.
A potential secondary effect, due to the moderately higher critical densities for the $v>1$ states than those of the $v=1$ states, is that a bottleneck can develop, such that the populations build up in the $v=1$ states.

Another effect of increasing thermalization is that
the relative column densities of molecules at high $J$ increase across all $v$ levels.  This is illustrated on the excitation diagram as the ``rotational temperature'' $T_{\rm rot}$, the excitation temperature derived from the populations of different rotational levels in the same vibration state,
increasing (i.e. each rotation ladder becomes more vertically compressed).   Panel (c) in Figure~\ref{fig:collisions} depicts the effect.
One possible explanation involves the competition between downward radiative and collisional transitions.
Downward radiative transitions have probabilities that strongly favor larger differences in energy ($\Delta E_{ul}$), selectively enhancing transitions to the lowest-$J$ levels, while downward collisional transitions do not strongly favor larger over smaller energy differences.
In a purely radiative cascade, the $\Delta J = -2$ transitions are favored over $\Delta J = 0$~or~$+2$, so $J$ tends to decrease during the cascade.
As gas density or temperature increases, a larger fraction of the rovibrational transitions will be collisionally induced, diluting the tendency for $J$ to decrease, leading to larger populations in levels at high $J$.
Another possible explanation is that this effect has little to do with collisions and
instead is simply a byproduct of how the level populations respond to increasing UV excitation rates compared to radiative and collisional de-excitation rates \citep[e.g. as argued in][]{draine96}.
A third possible explanation, previously discussed in K17,
relies on the fact that the majority of the \htwo{} in a PDR resides in the pure rotational states ($v=0$).
If the $v=0$ levels at high-$J$  are thermalized, their relative populations will rise with the kinetic temperature of the gas.
During UV excitation and the subsequent radiative cascade,
quantum selection rules allow transitions between all values of $\Delta v$ but restrict the value of $\Delta J$ to 0 or $\pm 2$.  Consequently, a molecule radiatively transitioning from $v=0$ to an excited electronic state and cascading back to the electronic ground state can take on any $v$ level but retains some ``memory'' of its original rotational ($J$) quantum number. 
For example, if an \htwo{} molecule is in a high $J$ level of an excited vibrational state, 
it cannot quickly lose this rotational 
energy in a singe radiative transition,
but must do so in a stepwise fashion, through multiple transitions among relatively high-$J$ levels, before reaching the $v=0$ state.  If the populations in the high-$J$ levels of $v=0$ are large, the imprint of the process that populates the $v \ge 1$ levels (through absorption of UV photons) is retained, enhancing the populations of high-$J$ levels in all $v$ states.
We have found that raising the density in PDR models with a fixed temperature results in higher rotation temperatures.
This favors the first explanation where increasing collision rates raises the fraction of collisionally induced vs. radiative transitions, which are more indifferent to changes in $J$.
The latter two explanations perhaps act to amplify the effect.

\afterpage{
\movetabledown=2cm
\begin{rotatetable}
\begin{deluxetable}{cccccccccccccc}
\centerwidetable
\tablecolumns{14}
\tabletypesize{\footnotesize}
\tablecaption{\htwo{} Rovibrational Level Column Densities ($N_u$) for Our PDRs and Comparison to Pure UV-excited Cloudy model ($N_m$)}\label{tab:sf-pdr-coldens}
\tablehead{\colhead{\htwo{} Line ID} & \colhead{$\lambda$ ($\mu$m)} & \colhead{$v_u$} & \colhead{$J_u$} & \multicolumn2c{S 140} & \multicolumn2c{NGC 2023} & \multicolumn2c{IC 63} & \multicolumn2c{Horsehead Nebula} & \multicolumn2c{Orion Bar}\\ \colhead{} & \colhead{} & \colhead{} & \colhead{} & \colhead{$\ln\left({N_u \over g_u}/{N_r \over g_r}\right)$} & \colhead{$N_u/N_m$} & \colhead{$\ln\left({N_u \over g_u}/{N_r \over g_r}\right)$} & \colhead{$N_u/N_m$} & \colhead{$\ln\left({N_u \over g_u}/{N_r \over g_r}\right)$} & \colhead{$N_u/N_m$} & \colhead{$\ln\left({N_u \over g_u}/{N_r \over g_r}\right)$} & \colhead{$N_u/N_m$} & \colhead{$\ln\left({N_u \over g_u}/{N_r \over g_r}\right)$} & \colhead{$N_u/N_m$}}
\startdata
1--0 Q(1) & 2.406592 & 1 & 1 &  $2.520^{+0.050}_{-0.053}$  & 1.94 & $3.117^{+0.075}_{-0.081}$  & 3.53 & $2.703^{+0.070}_{-0.076}$  & 2.33 & $2.714^{+0.040}_{-0.041}$  & 2.36 & $4.198^{+0.084}_{-0.092}$  & 10.40 \\
1--0 Q(2) & 2.413439 & 1 & 2 &  $2.167^{+0.098}_{-0.108}$  & 1.49 & $2.627^{+0.103}_{-0.114}$  & 2.35 & $2.303^{+0.114}_{-0.129}$  & 1.70 & $2.323^{+0.086}_{-0.095}$  & 1.74 & $3.685^{+0.085}_{-0.093}$  & 6.78 \\
1--0 S(0) & 2.223290 & 1 & 2 &  $2.166^{+0.092}_{-0.102}$  & 1.49 & $2.625^{+0.098}_{-0.108}$  & 2.35 & $2.333^{+0.110}_{-0.124}$  & 1.76 & $2.405^{+0.080}_{-0.087}$  & 1.89 & $3.729^{+0.078}_{-0.085}$  & 7.09 \\
1--0 S(1) & 2.121834 & 1 & 3 &  $1.607^{+0.045}_{-0.048}$  & 1.83 & $1.974^{+0.069}_{-0.074}$  & 2.65 & $1.754^{+0.067}_{-0.072}$  & 2.12 & $1.694^{+0.035}_{-0.036}$  & 2.00 & $3.010^{+0.074}_{-0.080}$  & 7.45 \\
1--0 Q(3) & 2.423730 & 1 & 3 &  $1.498^{+0.050}_{-0.053}$  & 1.66 & $1.856^{+0.075}_{-0.082}$  & 2.37 & $1.666^{+0.071}_{-0.076}$  & 1.96 & $1.689^{+0.040}_{-0.042}$  & 2.01 & $2.946^{+0.085}_{-0.093}$  & 7.06 \\
1--0 Q(4) & 2.437489 & 1 & 4 &  $0.931^{+0.098}_{-0.109}$  & 1.48 & $1.384^{+0.103}_{-0.115}$  & 2.33 & $1.105^{+0.114}_{-0.129}$  & 1.76 & $1.034^{+0.088}_{-0.096}$  & 1.64 & $2.429^{+0.086}_{-0.094}$  & 6.61 \\
1--0 S(2) & 2.033758 & 1 & 4 &  $0.915^{+0.086}_{-0.094}$  & 1.42 & $1.390^{+0.092}_{-0.101}$  & 2.28 & --  & --  &   $1.017^{+0.072}_{-0.078}$  & 1.57 & $2.389^{+0.071}_{-0.076}$  & 6.19 \\
1--0 S(3) & 1.957559 & 1 & 5 &  $0.453^{+0.042}_{-0.044}$  & 2.00 & $0.978^{+0.065}_{-0.070}$  & 3.38 & $0.751^{+0.064}_{-0.068}$  & 2.70 & $0.529^{+0.032}_{-0.033}$  & 2.16 & $1.740^{+0.067}_{-0.072}$  & 7.25 \\
1--0 Q(6) & 2.475559 & 1 & 6 &  $-0.406^{+0.101}_{-0.113}$  & 1.27 & $0.126^{+0.106}_{-0.119}$  & 2.16 & $-0.031^{+0.118}_{-0.134}$  & 1.84 & $-0.142^{+0.121}_{-0.138}$  & 1.65 & --  & --  \\ 
1--0 S(6) & 1.788050 & 1 & 8 &  --  & --  &   --  & --  &   --  & --  &   --  & --  &   $0.161^{+0.059}_{-0.063}$  & 2.71 \\
1--0 S(7) & 1.747955 & 1 & 9 &  $-1.755^{+0.038}_{-0.039}$  & 0.43 & $-1.002^{+0.059}_{-0.062}$  & 0.91 & $-1.208^{+0.060}_{-0.064}$  & 0.74 & $-1.923^{+0.039}_{-0.041}$  & 0.36 & $-0.054^{+0.057}_{-0.060}$  & 2.34 \\
1--0 S(8) & 1.714738 & 1 & 10 &  $-2.586^{+0.089}_{-0.098}$  & 0.22 & $-1.603^{+0.080}_{-0.087}$  & 0.58 & $-1.916^{+0.110}_{-0.123}$  & 0.43 & --  & --  &   $-0.473^{+0.056}_{-0.060}$  & 1.81 \\
1--0 S(9) & 1.687761 & 1 & 11 &  $-2.509^{+0.037}_{-0.039}$  & 0.32 & $-1.728^{+0.057}_{-0.060}$  & 0.70 & $-1.878^{+0.059}_{-0.063}$  & 0.60 & $-2.193^{+0.036}_{-0.038}$  & 0.44 & $-0.837^{+0.054}_{-0.057}$  & 1.71 \\
1--0 S(10) & 1.666475 & 1 & 12 &  $-3.228^{+0.110}_{-0.123}$  & 0.24 & $-2.123^{+0.081}_{-0.088}$  & 0.72 & $-2.587^{+0.138}_{-0.160}$  & 0.45 & --  & --  &   $-1.305^{+0.058}_{-0.062}$  & 1.63 \\
1--0 S(11) & 1.650413 & 1 & 13 &  --  & --  &   $-2.528^{+0.059}_{-0.063}$  & 0.82 & $-2.429^{+0.074}_{-0.079}$  & 0.91 & --  & --  &   $-1.931^{+0.057}_{-0.061}$  & 1.49 \\
\enddata
\tablecomments{For each PDR, all column densities are normalized to the reference value, $N_r$, derived from 4--2 O(3) line.\\
Undetected lines are marked with a dash.\\
(This table is available in its entirety in machine-readable form in the arXiv source.)}
\end{deluxetable}
\end{rotatetable}

\clearpage
}

\subsection{The Ortho-to-para Ratio}\label{sect:measure-ortho-to-para}

The \htwo{} molecular wave function allows for only odd or even levels of $J$ 
depending on whether the proton spins are aligned or opposite to each other.  Since  $\Delta J = 0$ or $\pm 2$ for radiative transitions,  these two distinct spin isomers  have essentially independent sets of rovibrational levels and are known as ortho-\htwo{} (spins aligned, odd $J$ levels) and para-\htwo{} (spins anti-aligned, even $J$ levels).  
In thermal equilibrium, the statistical weights for nuclear spin give an ortho-to-para abundance ratio (O/P) equal to 3.
Since radiative transitions disallow ortho-to-para conversion, the intrinsic O/P is set or modified only during \htwo{} formation, collisions between \htwo{} and other species such as H or H$^+$, or selective dissociation of either ortho- or para-\htwo{}.
\cite{sternberg99} point out that while the \textit{intrinsic} O/P in a PDR for \htwo{} in the ground ($v=0$) state is expected to be 3, if there is enough \htwo{} that 
the Lyman and Werner bands become optically thick, optical depth effects cause FUV photons to be differentially absorbed by ortho- and para-\htwo{} leading to an \textit{observed} O/P of $\sim1.7$ for \htwo{} excited by UV radiation into $v > 0$.  
Collisions can drive the observed O/P for these levels toward the thermal limit of 3. 
Consequently, the observed O/P might act as a probe for the collisional modification of the UV-excited \htwo{} spectrum in a PDR, independent of the other collisional effects on the rovibrational level populations, where the degree of collisional modification drives the value from 1.7 to 3.

We leverage the large number of \htwo{} rovibrational lines detected in each PDR to measure their observed O/P.
Our motivation is twofold.  First, we will correct the O/P so that the ortho- and para-\htwo{} rovibrational levels form a single continuous sequence for each rotation ladder on the excitation diagram.
Second, we aim to use the observed O/P as a probe of the collisional modification of the \htwo{} spectrum.
On an excitation diagram, we plot the level populations as $\ln\left(N_u/g_u\right)$ where
the statistical weight for the upper state of a transition for ortho-\htwo{} is $g_u^{\mbox{\tiny ortho}} = 3 (2 J_u + 1)$ and for para-\htwo{} is  $g_u^{\mbox{\tiny para}} = 2 J _u + 1$.
If the observed O/P is not the thermal value of 3, there will be some vertical distance on the excitation diagram separating the ortho and para levels in a given rotation ladder.
We measure the observed O/P by cross-correlating the ortho and para level populations over a range of possible O/P to find the value that results in the least vertical separation between the ortho and para levels. For each step in this cross-correlation, we apply an O/P correction to the ortho level populations and then linearly interpolate them for each rotation ladder in $\ln\left(N_u/g_u\right)$ space.  We then calculate the vertical separation between the para levels and interpolated ortho levels in each rotation ladder.  The squares of the separation distances are summed to calculate the Pearson $\chi^2$ statistic at each cross-correlation step, with the minimum $\chi^2$ giving the best-fit observed O/P.  We estimate the uncertainty from the standard deviation of 1000 bootstrapped samples of the cross-correlation.  The best-fit observed O/P and estimated uncertainties for each PDR are reported in Table~\ref{tab:sf-pdrs}, and used to correct all the ortho level populations in each PDR (Table~\ref{tab:sf-pdr-coldens}) to the intrinsic O/P of 3.

\section{Results}

\subsection{Comparison Among the Observed PDRs} \label{sect:pdr-comparison}

\begin{figure}
\begin{center}
\includegraphics[width=0.85\textwidth]{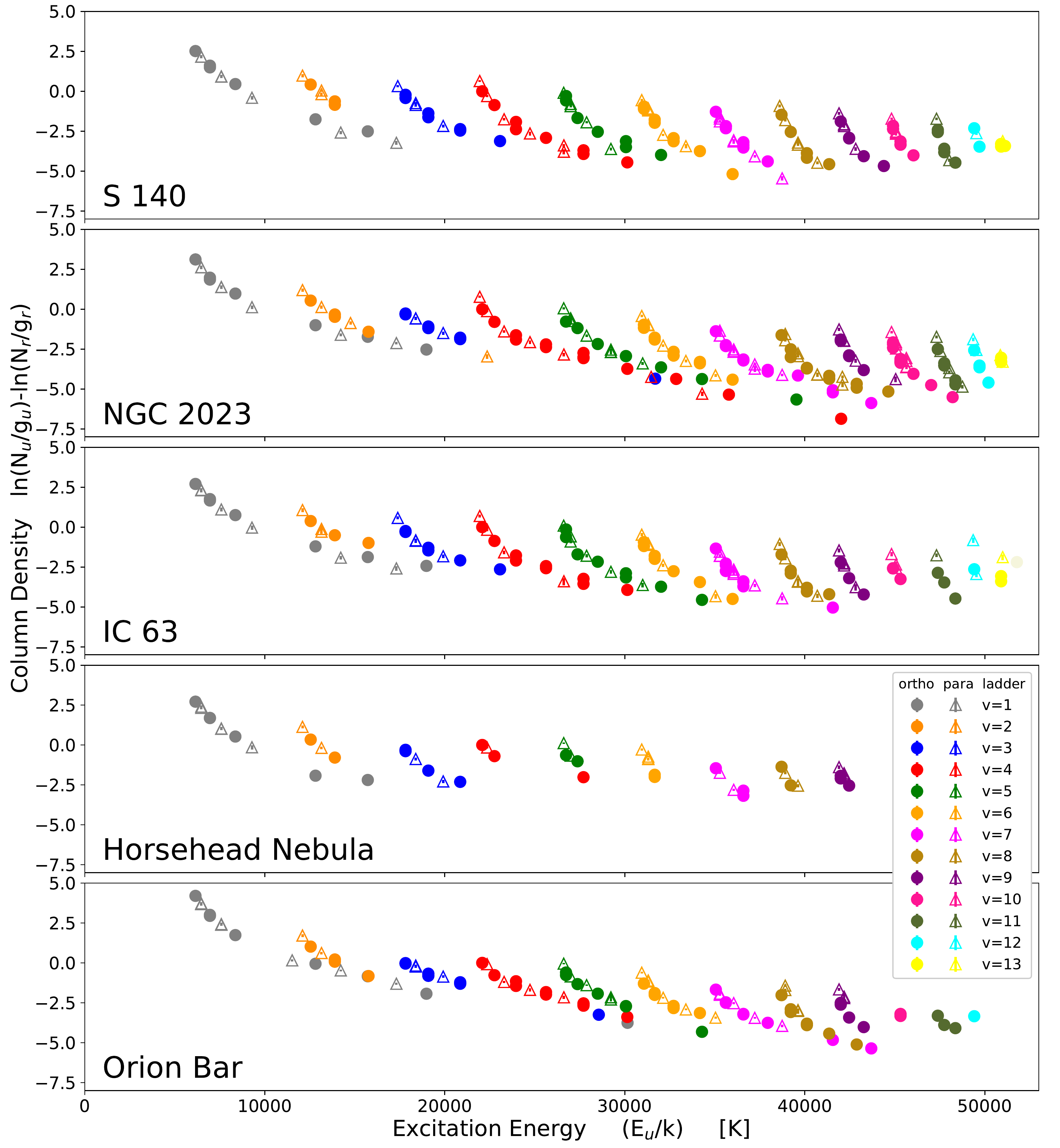}
\end{center}
\caption{Excitation diagrams showing the \htwo{} rovibrational level populations derived from our IGRINS observations of \htwo{} emission lines in each of the PDRs in this survey.  The column densities are given in Table~\ref{tab:sf-pdr-coldens}.
}
\label{fig:sf-pdr-excitation-diagrams}
\end{figure}

\begin{figure}
\begin{center}
\includegraphics[width=0.65\textwidth]{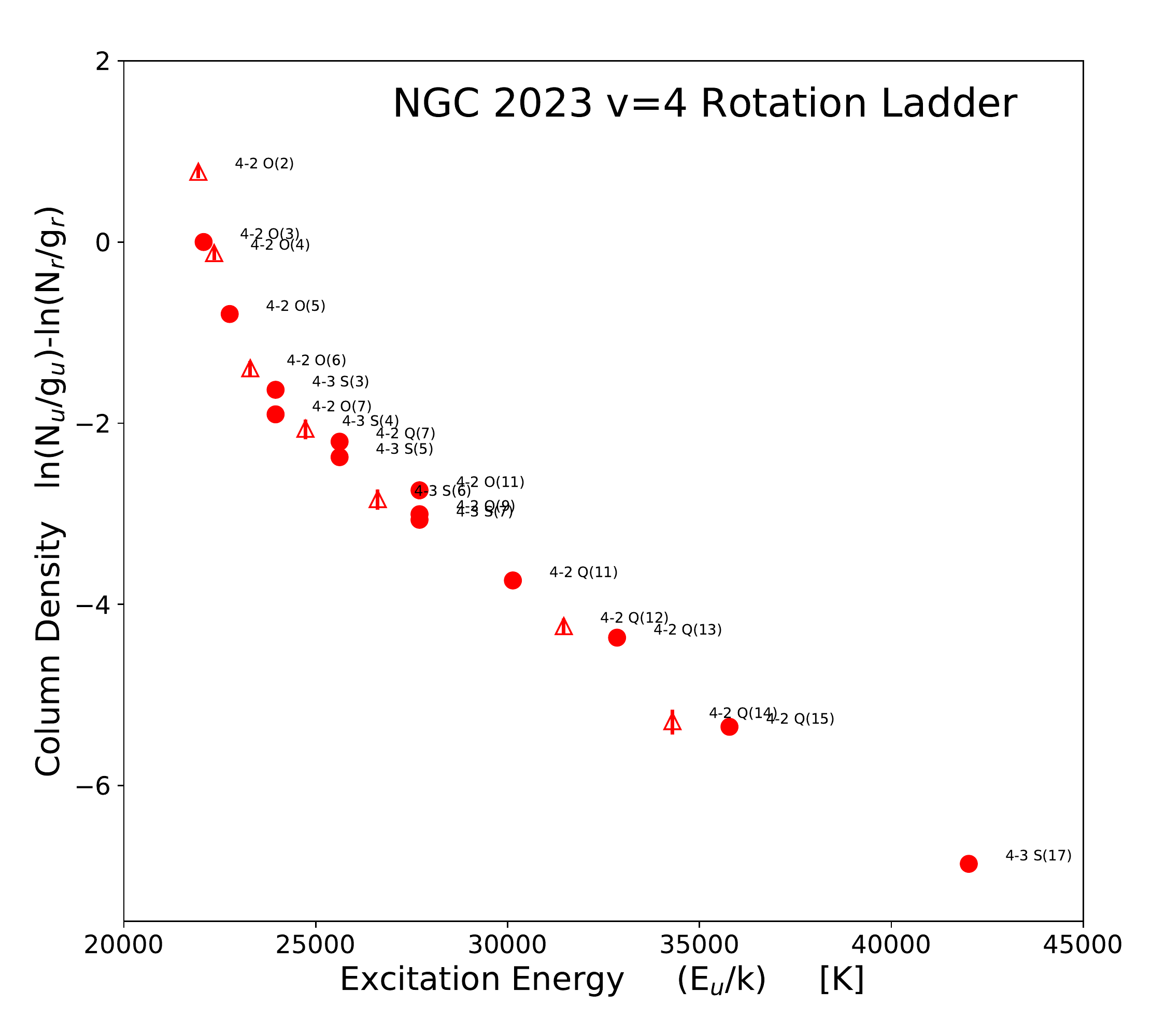}
\end{center}
\caption{Excitation diagram showing the \htwo{} rovibrational level populations of the $v=4$ rotational ladder in NGC 2023.  Circles denote ortho transitions, and triangles denote para transitions.
}
\label{fig:ngc2023-v4}
\end{figure}

Figure~\ref{fig:sf-pdr-excitation-diagrams} shows the excitation diagrams for all the PDRs in our survey.   
The extremely high signal-to-noise ratio (S/N) spectrum we obtained for NGC 2023 is especially impressive, with over 170 transitions detected with S/N~$> 3$.  The $v=4$ rotational ladder, displayed in Figure~\ref{fig:ngc2023-v4}, shows many transitions up to 4--3~S(17), which arises from $v=4$, $J = 19$ at an excitation energy of $E_u/k = 42,\!019$~K.
Qualitatively, the level populations seen in Table~\ref{tab:sf-pdr-coldens} and Figure~\ref{fig:sf-pdr-excitation-diagrams} appear similar across all the PDRs in our survey, showing the same 
nonthermal distribution
indicative of UV excitation.  This similarity is expected since the spectrum of UV-excited \htwo{} depends mainly on known physical constants and is fairly invariant to the conditions within the gas \citep{black87, sternberg89, shaw05}.

There are, however, subtle differences among the excitation diagrams of these PDRs, which encode information 
about conditions in the respective sources.
The nonthermal morphology is more prominent in S~140, NGC~2023, IC~63, and the Horsehead Nebula than in the Orion Bar.
In the particularly well-observed Orion Bar PDR, the jumps in column density at fixed $E_u$ from one $v$ level to the next are smaller, bringing the overall trend closer to a monotonic decline.  
For example, the $\ln(N_u / g_u)$ for $v=2, J=3$ (determined from the 2--1~S(1) transition) compared to $v=1, J=9$ (determined from the 1--0~S(7) transition), which share similar excitation energies near $E_u/k \sim 12,\!700$~K, show a difference 
in $\ln(N_u/g_u)$ space
of $\sim 1.1$ for the Orion Bar vs. a larger difference of $\sim 2.2$ for S~140.

Studies of the \htwo{} line emitting region in the Orion Bar
by \citet{parmar91}, \citet{allers05}, \citet{shaw09}, and K17 find temperatures of $250$--$1000$~K from the \htwo{} transitions.
A high temperature and/or density
increases the rate of collisions, which increases the collisional modification of UV-excited \htwo{} rovibrational level populations as discussed in \S~\ref{sec:h2levelpops}.
Steady-state PDR models underestimate the temperature at the \hi{}/\htwo{} dissociation interface.
Prior investigations suggest that this underestimation results from
either a clumpy medium \citep{burton90, parmar91, meixner93, andree-labsch14}, an additional heating mechanism \citep{allers05, pellegrini07, pellegrini09, shaw09},
higher than previously estimated \htwo{} formation rates \citep{lebourlot12, joblin18}, 
 or time-dependent effects, which the steady-state models fail to account for
\citep{hollenbach95,  storzer98, goicoechea16, goicoechea17, wu18}.
The Orion Bar is the source with the
highest $n_H$ and $G_0$ in our sample of PDRs (see Table~\ref{tab:sf-pdrs}), so perhaps this temperature discrepancy at the \hi{}/\htwo{} dissociation interface between the models and observations is strongest in PDRs with the highest density and/or strongest UV fields.

\subsection{The \lineratio{} Line Ratio} \label{sec:sfr-pdrs-h2-line-ratio}

\begin{figure}
\begin{center}
\includegraphics[width=0.6\textwidth]{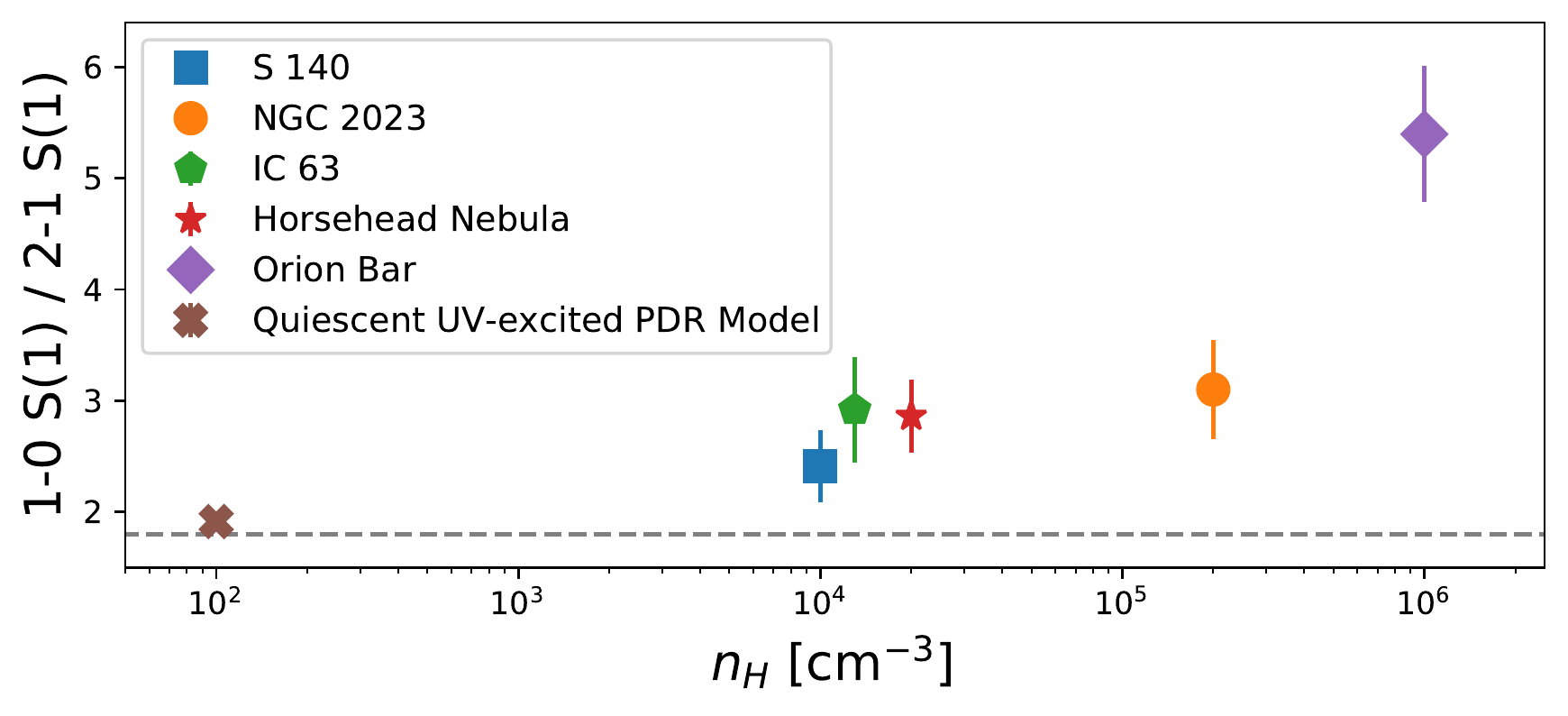} \\
\vspace{-0.15cm}
\includegraphics[width=0.6\textwidth]{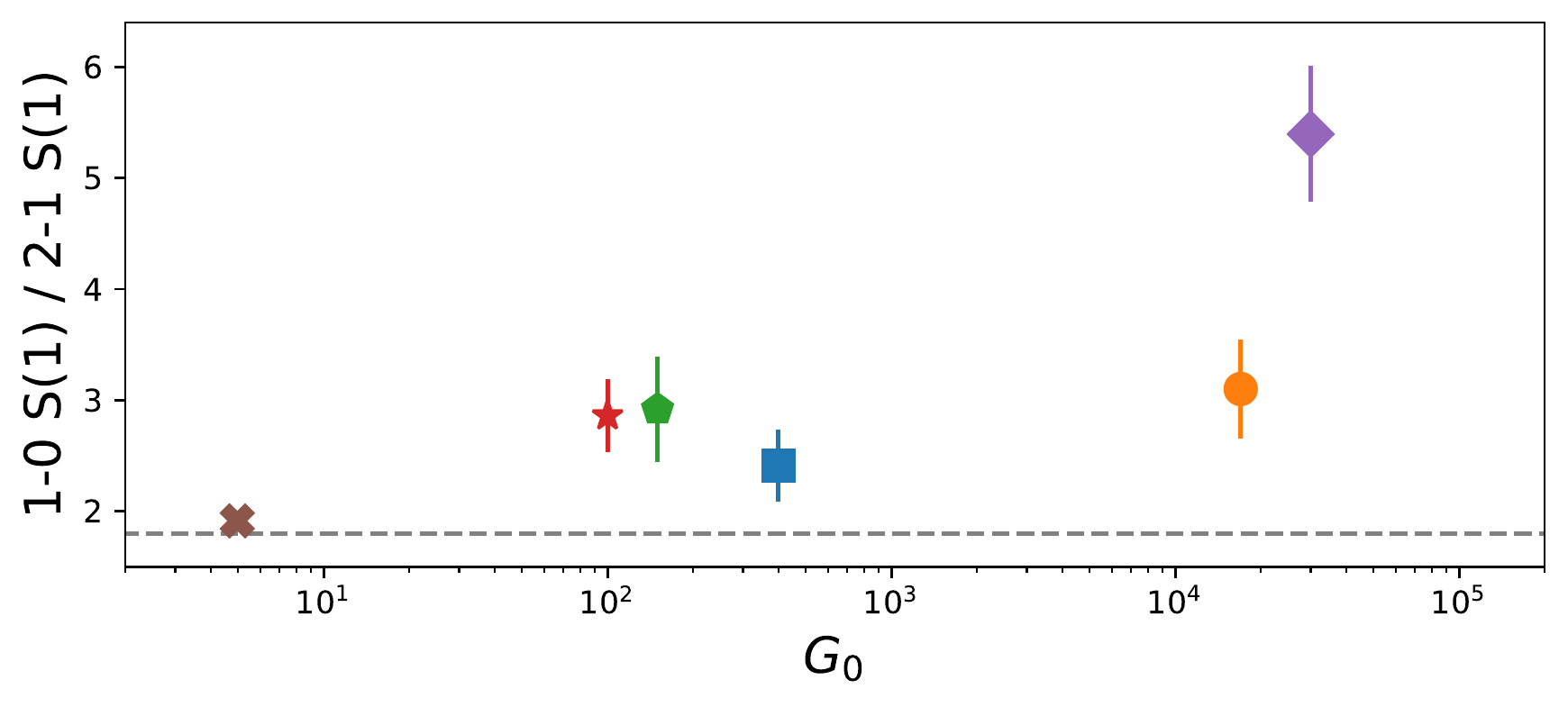} \\
\vspace{-0.15cm}
\includegraphics[width=0.6\textwidth]{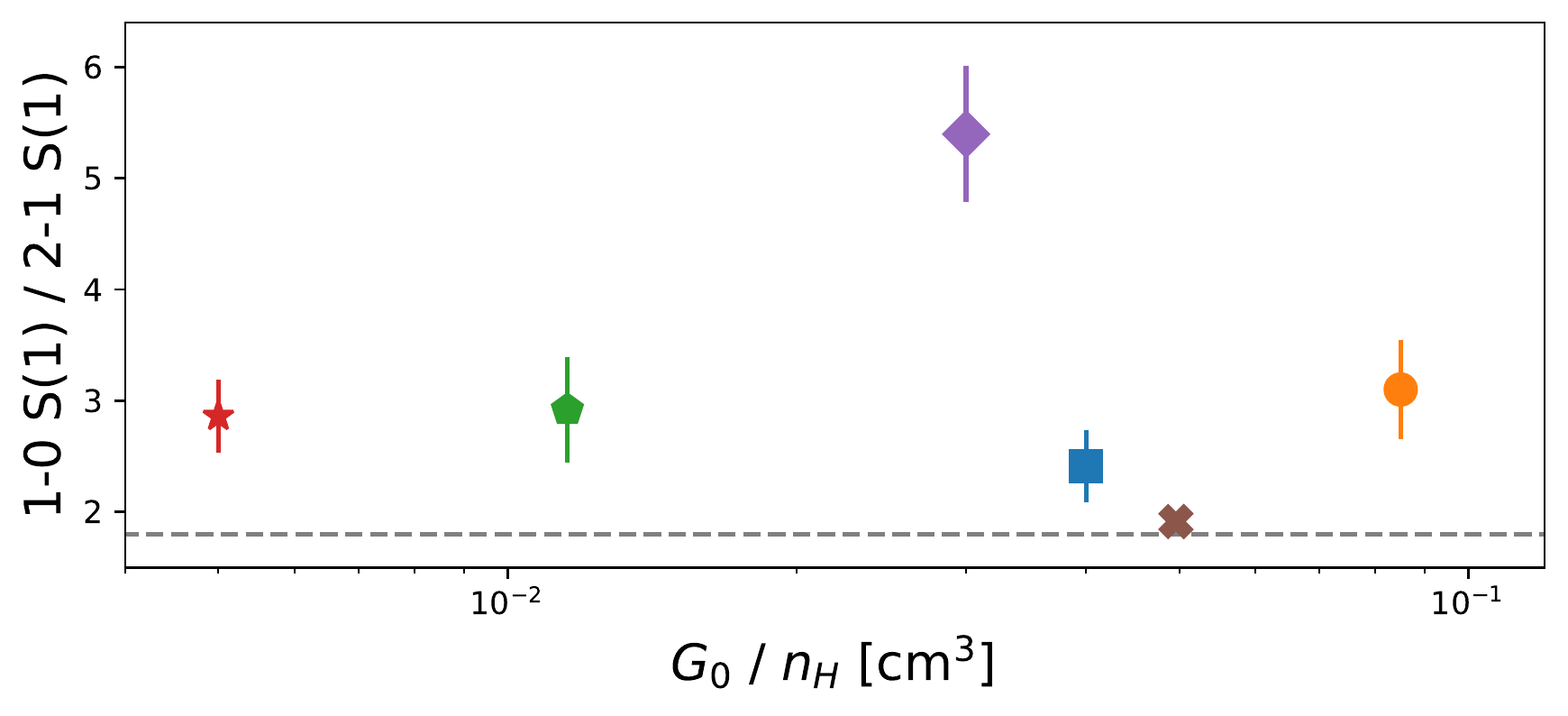}
\end{center}
\caption{
Semi-log plot of the \lineratio{} line ratio vs.
gas density $n_H$ (top), the UV field intensity $G_0$ (middle), and the ratio of the two $G_0/n_H$ (bottom).
The error bars show the 1$\sigma$ statistical uncertainty in the line ratio.
The \lineratio{} line ratio for unmodified UV excitation is $1.8$, which is represented by the horizontal gray dashed line, and the result from a quiescent low-density UV-excited PDR model is represented as
an $x$ symbol.
}
\label{fig:sf-pdr-ratios}
\end{figure}

The 1--0~S(1) and 2--1~S(1) lines are bright, easy to observe, and not near any major telluric absorption features.
The \lineratio{} line flux ratio has  been used historically as a convenient index to differentiate between shocked and UV-excited \htwo{} (e.g., \citealt{hayashi85, black87, burton92b}). 
For unmodified UV-excited \htwo{}, this line ratio is  $1.8$ \citep{black87}.
In \htwo{}-emitting regions with thermal level populations, such as shocks, the ratio can be very high. For example, \lineratio{} $\sim 200$ in a 1000~K gas.
Here {we repurpose it as a proxy for collisional modification of the \htwo{} rovibrational level populations.
All of our observed PDRs have \lineratio{} ratios $> 1.8$, indicating some degree of collisional modification. 
There are two ways in which collisions can drive up the \lineratio{} ratio:
(1) increasing the collisional de-excitation of the $v=2, J=3$ level relative to $v=1, J=3$, or (2) increasing the collisional excitation of the $v=1, J=3$ level relative to $v=2, J=3$.
These are the same processes depicted in panel (b) of Figure~\ref{fig:collisions}  (see \S~\ref{sec:h2levelpops})

In Table~\ref{tab:sf-pdrs} we report the value of \lineratio{} for each PDR along with values of $G_0$ and $n_H$ from the literature.  We also report the $G_0/n_H$ ratio, which parameterizes the number of UV photons per particle in the gas.  In Figure~\ref{fig:sf-pdr-ratios}, we plot \lineratio{} against $n_H$, $G_0$, and $G_0/n_H$.  We note that the uncertainties for \lineratio{} are high, so it is difficult to draw any definitive conclusions.  We also do not include uncertainties for $G_0$ or $n_H$ because most of these values from the literature are derived from model fits and lack formal uncertainties.
We expect the value of \lineratio{} to increase with increasing degree of collisional modification \citep{sternberg89, burton90, draine96}.  

We explore how the value of the \lineratio{} line ratio changes across our PDRs with respect to the UV field intensities $G_0$ and densities $n_H$, following discussions of the ratio in \citet{burton90} and \citet{draine96}.
\lineratio{} is mostly invariant over $n_H = 10^3$--$10^5$~cm$^{-3}$ and $G_0 = 10^2$--$2\times10^3$.  
S~140, the Horsehead Nebula, and IC~63  lie within this range of  $n_H$ and $G_0$ and have \lineratio{} values between 2.41 to 2.95. 
At densities above $n_H > 10^5$~cm$^{-3}$, the ratio becomes sensitive to both $n_H$ and $G_0$, and increases with both, due to increasing the collisional de-excitation of the $v=2, J=3$ level relative to $v=1, J=3$.
This trend in the \lineratio{} ratio with $n_H$ is most noticeable.  Our results are similar to \citet{luhman97b} who found that collisions start to affect the \htwo{} rovibrational level populations at $n_H > 5 \times 10^4$~cm$^{-3}$.
NGC 2023 lies beyond these ranges with $n_H = 2\times 10^5$~cm$^{-3}$ and $G_0 = 1.7 \times 10^4$, and has a slightly but not significantly higher \lineratio{} value of $3.10 \pm 0.44$. If the density is high ($n_H > 10^6$~cm$^{-3}$) and $G_0 > 10^4$, the gas becomes warm enough that collisional excitation starts to populate the $v=1, J=3$ levels, further raising the \lineratio{} ratio. 
The Orion Bar, where $G_0 = 3\times10^4$ and $n_H = 10^6$~cm$^{-3}$, exhibits the largest \lineratio{} value of $5.40 \pm 0.61$. 
Some models predict an even higher \lineratio{}.  For example, the best-fit isobaric model of the Orion Bar in \citet{joblin18}, has a gas pressure of $2.8 \times 10^8$~K~cm$^{-3}$ and recreates many of the observed features of the Orion Bar but  overestimates \lineratio{} by a factor of $\sim 10$.  This suggests that the pressure and density in the \htwo{} emitting gas is lower than in this model.   As discussed in $\S$~\ref{sect:pdr-comparison}, the \htwo{} emission in the Orion Bar has proven difficult to properly reproduce with steady-state models, and even the model of \citet{joblin18} which uses the latest treatments for \htwo{} formation \citep{lebourlot12},  likely requires additional mechanisms \citep[e.g. time-dependent effects;][]{goicoechea16, goicoechea17} to fully model the Orion Bar's \htwo{} emission.

\subsection{Comparison to a Quiescent UV-excited PDR Model} \label{sec:compare-to-model}

\begin{figure}
\begin{center}
\includegraphics[width=0.85\textwidth]{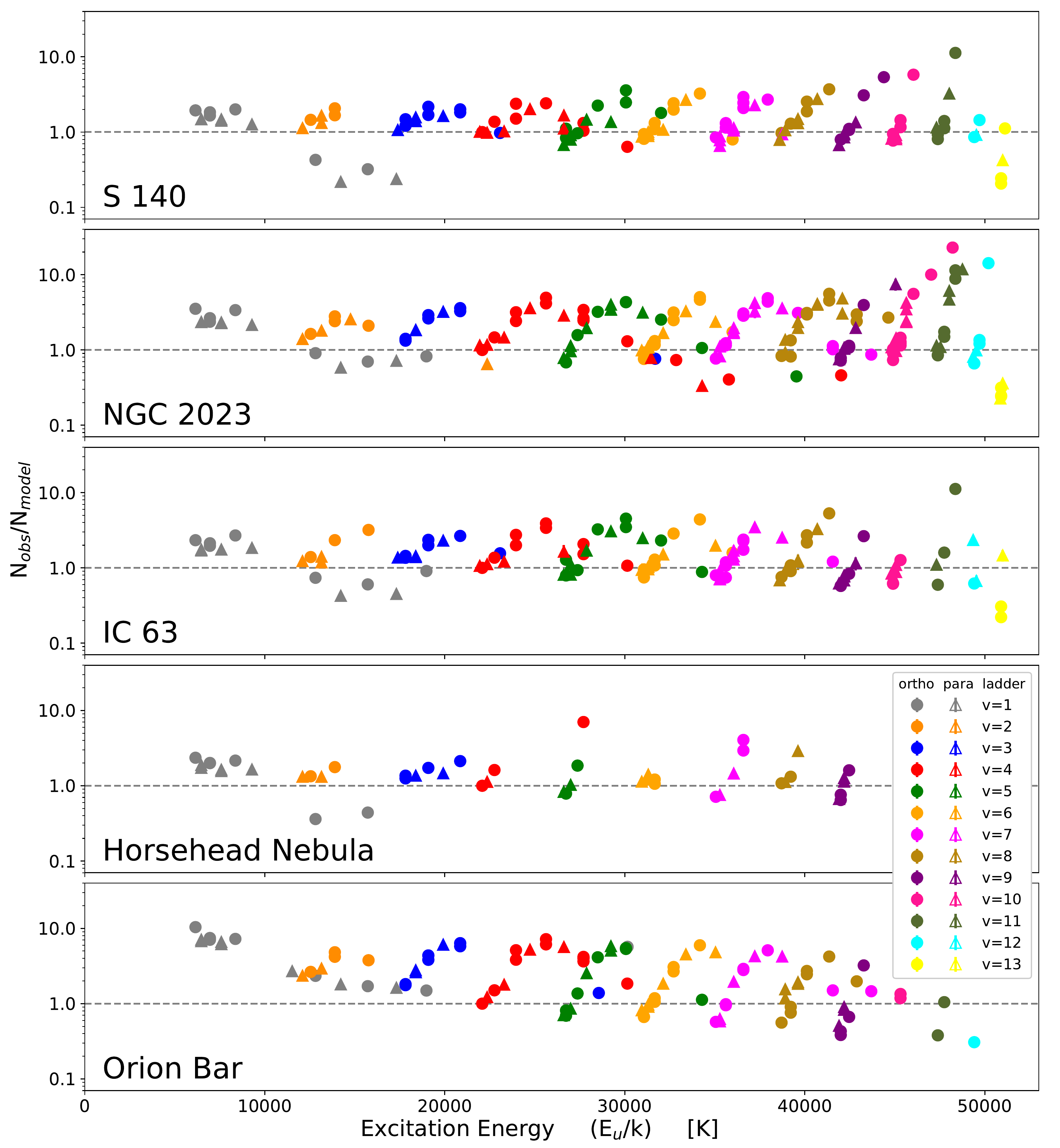}
\end{center}
\caption{Semi-log plots of the ratios of the observed rovibrational level populations in our data ($N_{\rm obs}$) to the  quiescent UV-excited Cloudy model  ($N_{\rm model}$). 
The horizontal dashed line represents a ratio of unity where the data and model are in perfect agreement.  Deviations from unity show where the level populations are not exactly matched by the model predictions.}
\label{fig:sf-pdr-pure-uv-ratios}
\end{figure}

An alternative method for quantifying the effects
of collisional modification of UV-excited \htwo{} level populations that takes full advantage of the rich NIR emission spectrum is to compare all the observed level populations to a Cloudy model \citep{cloudy13} of a quiescent, low-density PDR.
We ran a Cloudy (version C13.03) model with a constant density of $10^2$ cm$^{-3}$, no cosmic rays, a 27,000~K blackbody illuminating star with a luminosity of $L/L_\odot = 10^4$, and the cloud face 5 pc from the illuminating star.  We normalize the model to the column density of the $v=4$, $J=1$ level as we have done for our PDR sample, so both the model and observations are normalized in the same manner.
Table~\ref{tab:sf-pdr-coldens} and Figure~\ref{fig:sf-pdr-pure-uv-ratios}  present the ratios of the observed level populations for our PDRs to  the quiescent UV-excited model.

For this model, $G_0 / n_H = 4.96 \times 10^{-2}$, and the predicted \lineratio{} ratio is $1.914$, close to the expected value of $1.8$ for unmodified UV-excited \htwo{}. 
As can be seen in Figure \ref{fig:sf-pdr-pure-uv-ratios}, the effects of collisions on a purely UV-excited \htwo{} spectrum we described in \S~\ref{sec:h2levelpops} are readily apparent in all of our PDRs.
The measured low-$J$, $v=1$ level populations in our PDRs are consistently higher than those predicted by the model, and the low-$J$, $v>1$ levels are consistently lower.  
These differences are most evident in the Orion Bar, with deviations up to an order of magnitude from the model.
The dominant effects are that the collisional excitation increases the populations in the $v=1$ levels, and collisional de-excitation reduces the level populations for $v > 1$.
Also, the observed PDRs consistently have high-$J$ level populations in the $v>1$ states that are larger than those predicted by the model.  
The Orion Bar shows this effect for $v > 1$, and S~140, NGC~2023, and IC~63 show it for the $v > 2$ levels.  The situation is less clear in the 
Horsehead Nebula since we do not probe to sufficiently high $J$ levels in that source.  

We are unable to rule out the possibility that the observed differences between the model and PDR level populations are entirely due to collisions.
For example, our model uses the \citet{takahashi2001} prescription for \htwo{} formation pumping, which sets the rovibrational level populations for newly formed \htwo{}, but this prescription might not accurately predict the initial populations resulting from the formation process.
In the next section, we compare the PDRs to each other directly.

\pagebreak

\subsection{Comparison to S~140}

\begin{figure}
\begin{center}
\includegraphics[width=0.85\textwidth]{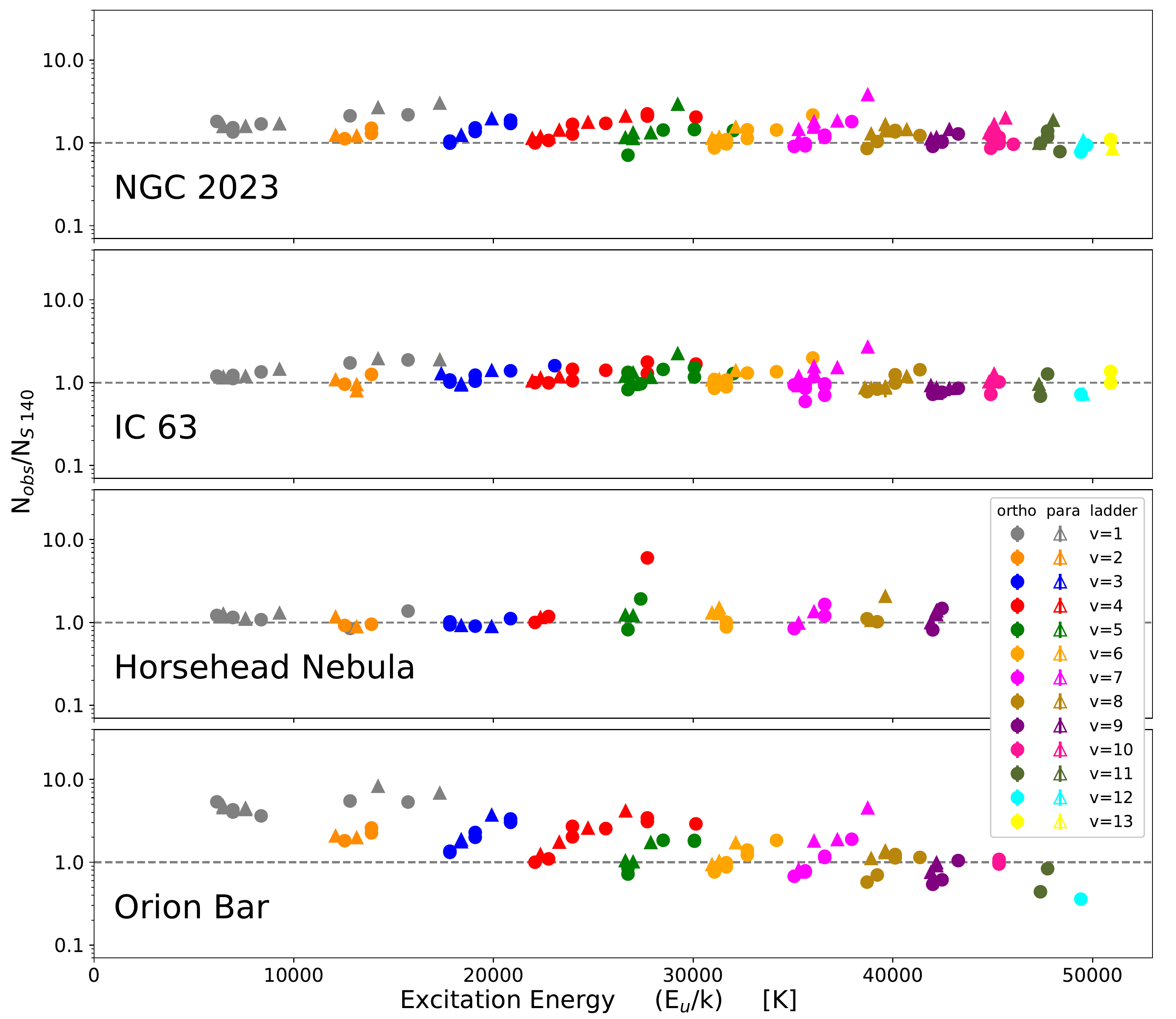}
\end{center}
\caption{Semi-log plots of the ratios of the observed rovibrational level populations in our data ($N_{\rm obs}$) to S 140  ($N_{\rm S\ 140}$).
The horizontal dashed line represents a ratio of unity where the data and S 140 are in perfect agreement.
}
\label{fig:sf-pdr-sh140-ratios}
\end{figure}

We also compare all the other PDRs in our survey to S~140.  This comparison avoids the assumptions and uncertainties in the PDR physics in the models by directly comparing the level populations.
We choose S~140 as our reference source because it is bright and probes a large number of \htwo{} rovibrational levels.  We can also test whether the S~140 \htwo{} spectrum shows the least collisional modification, as implied by its low 1--0~S(1)/2--1~S(1) ratio. 
For this comparison, we average all column densities derived for each $v$ and $J$ level in S~140 in cases where multiple transitions arising from the same upper level were observed. 
We plot the ratio of the observed level populations for all the PDRs to  S~140 in Figure~\ref{fig:sf-pdr-sh140-ratios}.

The results are similar to the comparison to the Cloudy model.
As we initially expected, most of the other PDRs show a greater degree of collisional modification in their rovibrational level populations than is found in S~140.  
In particular, the other PDRs show higher populations in the $v=1$ levels and in the high $J$ levels at all $v$, consistent with enhancements due to collisional effects as described in \S~\ref{sec:h2levelpops}.
The results are somewhat ambiguous for the Horsehead Nebula, due to the lower number of rovibrational levels detected in this relatively faint source.
As we saw earlier in the comparison to the model, the Orion Bar shows the greatest degree of collisional modification to its level populations.

\subsection{Ortho-to-Para Ratio Comparison}\label{sect:ortho-to-para-ratio-results}

\begin{figure}
\begin{center}
\includegraphics[width=0.6\textwidth]{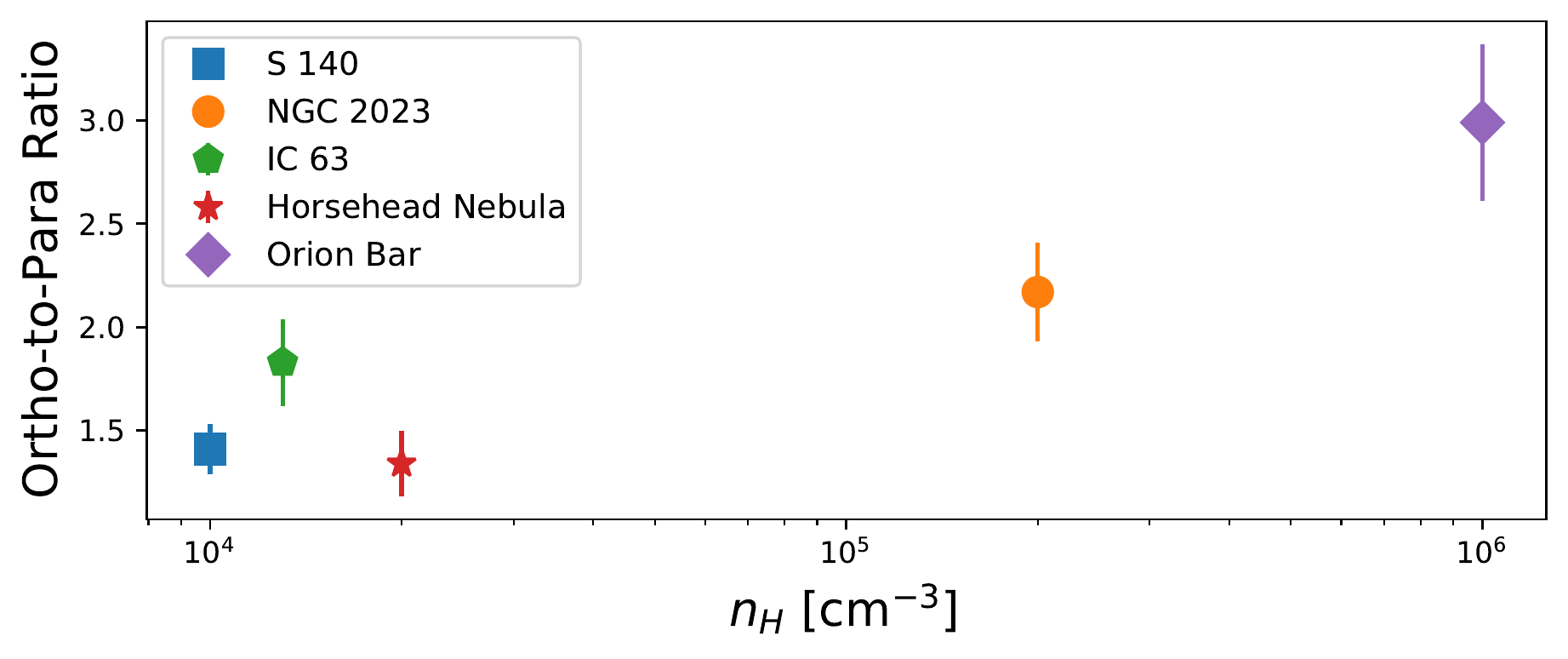} \\
\vspace{-0.15cm}
\includegraphics[width=0.6\textwidth]{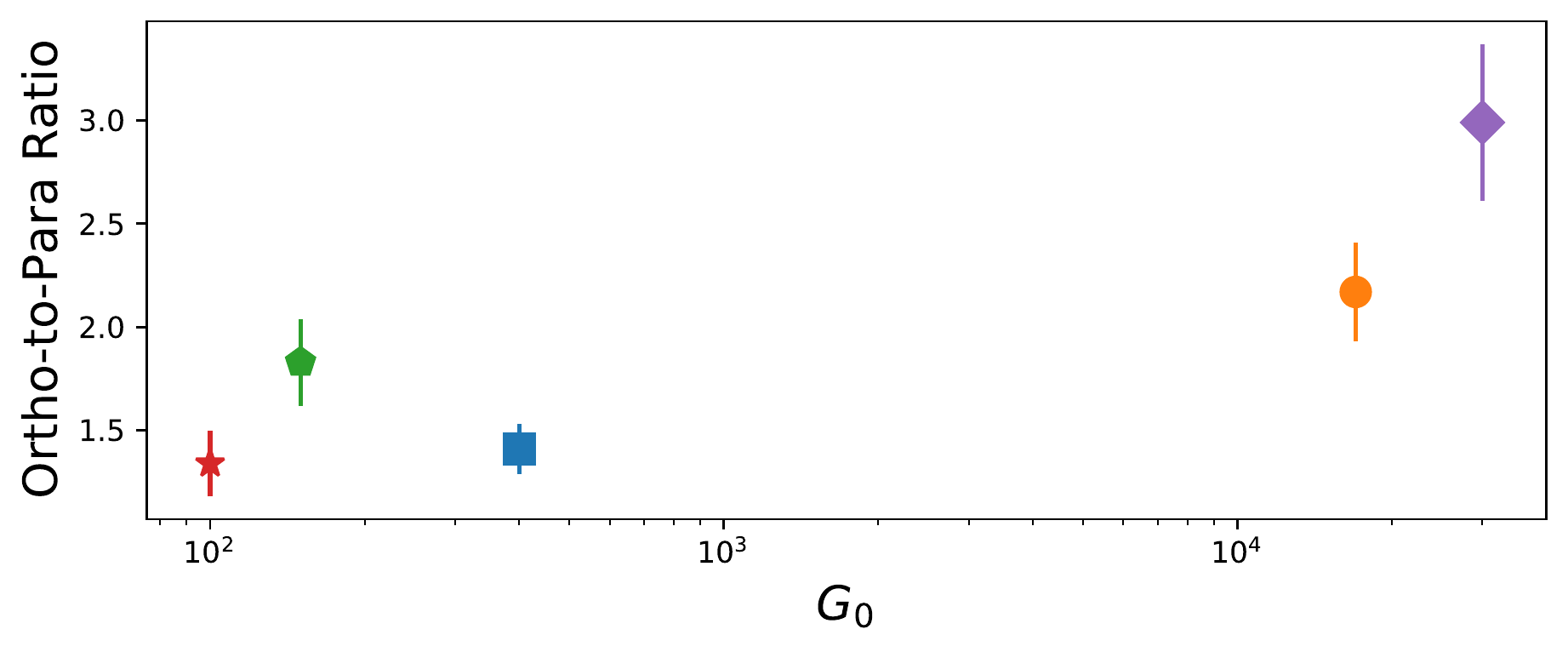} \\
\vspace{-0.15cm}
\includegraphics[width=0.6\textwidth]{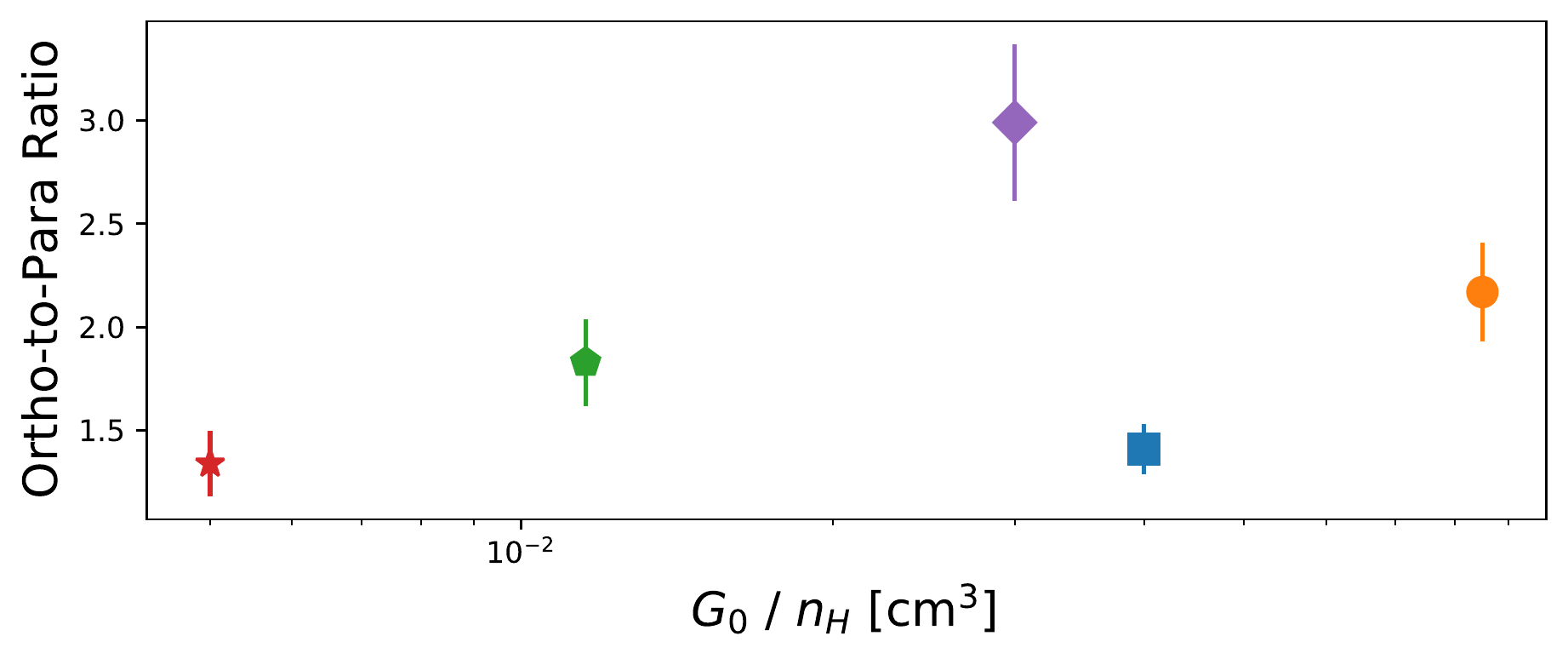}
\end{center}
\caption{
Semi-log plot of the best-fit observed ortho-to-para ratio (O/P) vs. the
gas density $n_H$ (top), the UV field intensity $G_0$ (middle), and the ratio of the two $G_0/n_H$ (bottom).
The error bars show the 1$\sigma$ statistical uncertainty in the ratio.
The expected O/P is $\sim1.7$ for UV-excited \htwo{} and approaches the thermal value of 3 with as the collisional modification of the of the UV-excited \htwo{} spectrum increases.
}
\label{fig:ortho-to-para-ratios}
\end{figure}

In \S{} \ref{sect:measure-ortho-to-para} we estimated the O/P ratio in each PDR and reported our results in Table~\ref{tab:sf-pdrs}.
The expected intrinsic O/P for \htwo{} in thermal equilibrium is 3.
In quiescent PDRs where the excitation is primarily caused by UV photons, the observed O/P for rovibrationally excited \htwo{} is $\sim 1.7$ due to optical depth effects of the FUV photons being differentially absorbed by ortho- and para-\htwo{} \citep{sternberg99}.

In Figure \ref{fig:ortho-to-para-ratios} we plot the O/P vs. $n_H$, $G_0$, and $G_0/n_H$. 
Our results for the O/P resemble our results for \lineratio{} where the effects of collisions start to become noticeable in NGC 2023 and become much more prominent in the Orion Bar. 
S~140, the Horsehead Nebula, and IC~63 have the lowest O/P of $1.41\pm0.12$, $1.34\pm0.16$, and $1.83\pm0.21$, respectively, near the expected value of $\sim 1.7$.
NGC~2023, with higher values of  $n_H$ and $G_0$ has a higher ratio of $2.17\pm0.24$.  The Orion Bar, with the highest values of $n_H$ and $G_0$, has the highest ratio of $2.99\pm0.38$. 

These results roughly follow the models of \citet{draine96} where O/P is low at when $n_H <  10^4$~cm$^{-3}$, but if $n_H >> 10^4$ the O/P will approach a value of 3 as $n_H$ increases. 
Several explanations for this trend have been suggested.    \citet{draine96} discuss how preferential self-shielding of rovibrationally excited ortho-\htwo{} can raise the observed O/P, even above the thermal limit of 3, although the Orion Bar's O/P of $2.99\pm0.38$ is indistinguishable from the thermal limit.
\citet{sternberg99} discuss the effects of collisional excitation and de-excitation.  Collisional excitation can raise the observed O/P in the lower-energy $v=1$ levels.  Since we fit all $v$ levels to calculate the O/P, collisional excitation of the lower-energy $v=1$ levels cannot fully explain our results. 
\citet{sternberg99} also discuss the possibility that collisional de-excitation during the radiative cascade can lead to ortho-para conversions and drive the observed O/P to the thermal value of 3.  This explanation is consistent with the other observed trends in the \htwo{} rovibrational level populations that we have attributed to collisions; although we cannot rule out the other effects or if unaccounted for mechanisms are affecting the observed O/P.

\section{Summary and Conclusions}

Dense molecular clouds respond to UV illumination from nearby hot stars by producing a rich spectrum of \htwo{} rovibrational lines in the near-IR.
We have conducted a survey of five PDRs in regions of high-mass star formation: S~140, NGC~2023, IC~63, the Horsehead Nebula, and the Orion Bar, that span a range of gas densities, UV field intensities, and illuminating star properties.  The PDRs were observed with single deep IGRINS pointings to maximize the S/N.  Each PDR displays a multitude of narrow \htwo{} lines (low velocity widths) from excited rovibrational energy levels with flux ratios clearly indicative of UV (fluorescent) excitation.
From these line fluxes, we calculate the detailed \htwo{} rovibrational level populations in each PDR.  The relative level populations reflect the combined effects of UV  excitation and radiative de-excitation along with collisional excitation and de-excitation.  As a result, they provide information about both the radiative environment and the physical conditions in the molecular gas at the 
PDR boundary.

We have found the following:

\begin{enumerate}
\item In all of the PDRs we observed, the lines are narrow ($< 10$~km~s$^{-1}$)  with internal motions of only a few kilometers per second and do not show any evidence for broad components that might result from interactions with stellar winds, clump-clump collisions, or shocks.  The kinematics point strongly to UV excitation (fluorescence) and radiative heating as the power behind the emission we observe.
\item Many lines were observed in each PDR, with the spectrum of NGC 2023 showing over 170 \htwo{} lines observed with S/N~$\ge 3$.
This spectrum includes transitions up to very  high $J$ levels, e.g. 4--3~S(17) which arise from the $v=4$, $J=19$ level at an excitation energy of $E_u/k = 42,\!019$ K.
\item All of the observed PDRs display values of the widely used diagnostic flux ratio of the \lineratio{} lines that are larger than the value for pure UV excitation, with the largest deviation seen for the Orion Bar,
the source with the highest value of  $n_H$ and $G_0$.
However, this simple index is insufficient to capture subtle differences in the populations of levels with higher $v$ and $J$ values for different PDRs.
\item
While the excitation diagrams for the PDRs all show
nonthermal distributions, as
one would expect from UV-excited \htwo{},
their level populations all
exhibit evidence for some degree of collisional modification relative to a Cloudy model of a pure UV-excited region.
This evidence is comprised of enhanced populations in the $v=1$ levels relative to those of the $v > 1$ levels
and of  the fact that
the $J$ levels within each $v$ ladder exhibit a pattern of systematically increasing population enhancements with larger $J$.  We discuss several possible mechanisms that could cause or contribute to these enhancements at high $J$.
\item Comparison of the five observed PDRs indicates that S~140 shows the least degree of collisional modification of a pure UV-excited spectrum in our sample, while the Orion Bar shows the greatest degree of modification.
The effect of the trend with $J$ discussed above is less apparent when comparing the other PDRs to S~140 than when comparing to the model.
\item 
Several of the PDRs in our sample show ortho-to-para ratios in excited rovibrational energy levels consistent with the value of $~1.7$ predicted by \citet{sternberg99} for pure UV-excited \htwo{}.
As $n_H$ increases,
these values approach the thermal equilibrium value of 3, with the Orion Bar's ratio being indistinguishable from the thermal value.  We discuss several possible explanations for this trend, including the effect of collisions during the radiative cascade.
\item Of the PDRs in our sample, the Orion Bar shows the greatest degree of collisional modification in its level populations relative to the model and the PDRs in our survey.
The Orion Bar has the highest 
$n_H$ and $G_0$ in our sample.
The \htwo{} rovibrational level populations in high density PDRs subject to strong UV fields, like the Orion Bar, are quantitatively different from the relative populations in lower-density PDRs illuminated by lower-intensity UV fields, showing greater effects from collisions.
\item Deep high-spectral resolution measurements of the \htwo{} emission in PDRs, such as those presented here, provide a valuable tool for probing the behavior of \htwo{} in UV radiation fields, a ubiquitous component of the interstellar medium of galaxies.  As illustrated in this survey, detailed comparisons of observed line ratios in different regions and with models can reveal dependencies on physical conditions and constrain critical aspects of molecular physical processes, which can be applied to the interpretation of future observations.
\end{enumerate}

\acknowledgments{
This work used the Immersion Grating Infrared Spectrometer (IGRINS) that was developed under a collaboration between the University of Texas at Austin and the Korea Astronomy and Space Science Institute (KASI) with the financial support of the Mt. Cuba Astronomical Foundation, of the US National Science Foundation under grants AST-1229522 and AST-1702267, of the McDonald Observatory of the University of Texas at Austin, of the Korean GMT Project of KASI, and Gemini Observatory.

We would like to acknowledge all the members of the IGRINS team whose work and support made these observations possible.

This paper includes data taken at The McDonald Observatory of The University of Texas at Austin.

H.L.D. received support from NSF grant AST-1715332.

We would like to thank Amiel Sternberg for his comments on an earlier draft of this paper.

We also would like to acknowledge the helpful comments and suggestions from the anonymous reviewer.

This work is based in part on observations made with the Spitzer Space Telescope, obtained from the NASA/ IPAC Infrared Science Archive, both of which are operated by the Jet Propulsion Laboratory, California Institute of Technology under a contract with the National Aeronautics and Space Administration.
}


\begin{thebibliography}{}
\expandafter\ifx\csname natexlab\endcsname\relax\def\natexlab#1{#1}\fi
\providecommand{\url}[1]{\href{#1}{#1}}
\providecommand{\dodoi}[1]{doi:~\href{http://doi.org/#1}{\nolinkurl{#1}}}
\providecommand{\doeprint}[1]{\href{http://ascl.net/#1}{\nolinkurl{http://ascl.net/#1}}}
\providecommand{\doarXiv}[1]{\href{https://arxiv.org/abs/#1}{\nolinkurl{https://arxiv.org/abs/#1}}}

\bibitem[{{Abergel} {et~al.}(2003){Abergel}, {Teyssier}, {Bernard},
  {Boulanger}, {Coulais}, {Fosse}, {Falgarone}, {Gerin}, {Perault}, {Puget},
  {Nordh}, {Olofsson}, {Huldtgren}, {Kaas}, {Andr{\'e}}, {Bontemps}, {Casali},
  {Cesarsky}, {Copet}, {Davies}, {Montmerle}, {Persi}, \&
  {Sibille}}]{abergel2003}
{Abergel}, A., {Teyssier}, D., {Bernard}, J.~P., {et~al.} 2003, \aap, 410, 577,
  \dodoi{10.1051/0004-6361:20030878}

\bibitem[{{Allers} {et~al.}(2005){Allers}, {Jaffe}, {Lacy}, {Draine}, \&
  {Richter}}]{allers05}
{Allers}, K.~N., {Jaffe}, D.~T., {Lacy}, J.~H., {Draine}, B.~T., \& {Richter},
  M.~J. 2005, \apj, 630, 368, \dodoi{10.1086/431919}

\bibitem[{{Andree-Labsch} {et~al.}(2017){Andree-Labsch}, {Ossenkopf-Okada}, \&
  {R{\"o}llig}}]{andree-labsch14}
{Andree-Labsch}, S., {Ossenkopf-Okada}, V., \& {R{\"o}llig}, M. 2017, \aap,
  598, A2, \dodoi{10.1051/0004-6361/201424287}

\bibitem[{{Andrews} {et~al.}(2018){Andrews}, {Peeters}, {Tielens}, \&
  {Okada}}]{andrews18}
{Andrews}, H., {Peeters}, E., {Tielens}, A.~G.~G.~M., \& {Okada}, Y. 2018,
  \aap, 619, A170, \dodoi{10.1051/0004-6361/201832808}

\bibitem[{{Black} \& {Dalgarno}(1976)}]{black76}
{Black}, J.~H., \& {Dalgarno}, A. 1976, \apj, 203, 132, \dodoi{10.1086/154055}

\bibitem[{{Black} \& {van Dishoeck}(1987)}]{black87}
{Black}, J.~H., \& {van Dishoeck}, E.~F. 1987, \apj, 322, 412,
  \dodoi{10.1086/165740}

\bibitem[{{Burton}(1992)}]{burton92b}
{Burton}, M.~G. 1992, Australian Journal of Physics, 45, 463,
  \dodoi{10.1071/PH920463}

\bibitem[{{Burton} {et~al.}(1990{\natexlab{a}}){Burton}, {Geballe}, {Brand}, \&
  {Moorhouse}}]{burton90b}
{Burton}, M.~G., {Geballe}, T.~R., {Brand}, P.~W.~J.~L., \& {Moorhouse}, A.
  1990{\natexlab{a}}, \apj, 352, 625, \dodoi{10.1086/168564}

\bibitem[{{Burton} {et~al.}(1990{\natexlab{b}}){Burton}, {Hollenbach}, \&
  {Tielens}}]{burton90}
{Burton}, M.~G., {Hollenbach}, D.~J., \& {Tielens}, A.~G.~G.~M.
  1990{\natexlab{b}}, \apj, 365, 620, \dodoi{10.1086/169516}

\bibitem[{{Burton} {et~al.}(1998){Burton}, {Howe}, {Geballe}, \&
  {Brand}}]{burton98}
{Burton}, M.~G., {Howe}, J.~E., {Geballe}, T.~R., \& {Brand}, P.~W.~J.~L. 1998,
  \pasa, 15, 194, \dodoi{10.1071/AS98194}

\bibitem[{{Caballero}(2008)}]{caballero08}
{Caballero}, J.~A. 2008, \mnras, 383, 750,
  \dodoi{10.1111/j.1365-2966.2007.12614.x}

\bibitem[{{Compi{\`e}gne} {et~al.}(2008){Compi{\`e}gne}, {Abergel},
  {Verstraete}, \& {Habart}}]{compiegne08}
{Compi{\`e}gne}, M., {Abergel}, A., {Verstraete}, L., \& {Habart}, E. 2008,
  \aap, 491, 797, \dodoi{10.1051/0004-6361:200809850}

\bibitem[{{Davis} {et~al.}(2003){Davis}, {Smith}, {Stern}, {Kerr}, \&
  {Chiar}}]{davis03}
{Davis}, C.~J., {Smith}, M.~D., {Stern}, L., {Kerr}, T.~H., \& {Chiar}, J.~E.
  2003, \mnras, 344, 262, \dodoi{10.1046/j.1365-8711.2003.06820.x}

\bibitem[{{Draine}(2003)}]{draine03}
{Draine}, B.~T. 2003, \araa, 41, 241,
  \dodoi{10.1146/annurev.astro.41.011802.094840}

\bibitem[{{Draine} \& {Bertoldi}(1996)}]{draine96}
{Draine}, B.~T., \& {Bertoldi}, F. 1996, \apj, 468, 269, \dodoi{10.1086/177689}

\bibitem[{{Fazio} {et~al.}(2004){Fazio}, {Hora}, {Allen}, {Ashby}, {Barmby},
  {Deutsch}, {Huang}, {Kleiner}, {Marengo}, {Megeath}, {Melnick}, {Pahre},
  {Patten}, {Polizotti}, {Smith}, {Taylor}, {Wang}, {Willner}, {Hoffmann},
  {Pipher}, {Forrest}, {McMurty}, {McCreight}, {McKelvey}, {McMurray}, {Koch},
  {Moseley}, {Arendt}, {Mentzell}, {Marx}, {Losch}, {Mayman}, {Eichhorn},
  {Krebs}, {Jhabvala}, {Gezari}, {Fixsen}, {Flores}, {Shakoorzadeh}, {Jungo},
  {Hakun}, {Workman}, {Karpati}, {Kichak}, {Whitley}, {Mann}, {Tollestrup},
  {Eisenhardt}, {Stern}, {Gorjian}, {Bhattacharya}, {Carey}, {Nelson},
  {Glaccum}, {Lacy}, {Lowrance}, {Laine}, {Reach}, {Stauffer}, {Surace},
  {Wilson}, {Wright}, {Hoffman}, {Domingo}, \& {Cohen}}]{fazio04}
{Fazio}, G.~G., {Hora}, J.~L., {Allen}, L.~E., {et~al.} 2004, \apjs, 154, 10,
  \dodoi{10.1086/422843}

\bibitem[{{Ferland} {et~al.}(2012){Ferland}, {Henney}, {O'Dell}, {Porter}, {van
  Hoof}, \& {Williams}}]{ferland12}
{Ferland}, G.~J., {Henney}, W.~J., {O'Dell}, C.~R., {et~al.} 2012, \apj, 757,
  79, \dodoi{10.1088/0004-637X/757/1/79}

\bibitem[{{Ferland} {et~al.}(2013){Ferland}, {Porter}, {van Hoof}, {Williams},
  {Abel}, {Lykins}, {Shaw}, {Henney}, \& {Stancil}}]{cloudy13}
{Ferland}, G.~J., {Porter}, R.~L., {van Hoof}, P.~A.~M., {et~al.} 2013, RMxAA,
  49, 137.
\newblock \url{https://ui.adsabs.harvard.edu/abs/2013RMxAA..49..137F}

\bibitem[{{Field} {et~al.}(1994){Field}, {Gerin}, {Leach}, {Lemaire}, {Pineau
  Des Forets}, {Rostas}, {Rouan}, \& {Simons}}]{field94}
{Field}, D., {Gerin}, M., {Leach}, S., {et~al.} 1994, \aap, 286, 909
\newblock \url{https://ui.adsabs.harvard.edu/abs/1994A%26A...286..909F}

\bibitem[{{Field} {et~al.}(1998){Field}, {Lemaire}, {Pineau des Forets},
  {Gerin}, {Leach}, {Rostas}, \& {Rouan}}]{field98}
{Field}, D., {Lemaire}, J.~L., {Pineau des Forets}, G., {et~al.} 1998, \aap,
  333, 280
\newblock \url{https://ui.adsabs.harvard.edu/abs/1998A%26A...333..280F}

\bibitem[{{Field} {et~al.}(1966){Field}, {Somerville}, \& {Dressler}}]{field66}
{Field}, G.~B., {Somerville}, W.~B., \& {Dressler}, K. 1966, \araa, 4, 207,
  \dodoi{10.1146/annurev.aa.04.090166.001231}

\bibitem[{{Fleming} {et~al.}(2010){Fleming}, {France}, {Lupu}, \&
  {McCandliss}}]{fleming10}
{Fleming}, B., {France}, K., {Lupu}, R.~E., \& {McCandliss}, S.~R. 2010, \apj,
  725, 159, \dodoi{10.1088/0004-637X/725/1/159}

\bibitem[{{France} {et~al.}(2005){France}, {Andersson}, {McCandliss}, \&
  {Feldman}}]{france05}
{France}, K., {Andersson}, B.-G., {McCandliss}, S.~R., \& {Feldman}, P.~D.
  2005, \apj, 628, 750, \dodoi{10.1086/430878}

\bibitem[{{Gaia Collaboration} {et~al.}(2018){Gaia Collaboration}, {Brown},
  {Vallenari}, {Prusti}, {de Bruijne}, {Babusiaux}, {Bailer-Jones}, {Biermann},
  {Evans}, {Eyer}, {Jansen}, {Jordi}, {Klioner}, {Lammers}, {Lindegren},
  {Luri}, {Mignard}, {Panem}, {Pourbaix}, {Randich}, {Sartoretti}, {Siddiqui},
  {Soubiran}, {van Leeuwen}, {Walton}, {Arenou}, {Bastian}, {Cropper},
  {Drimmel}, {Katz}, {Lattanzi}, {Bakker}, {Cacciari}, {Casta{\~n}eda},
  {Chaoul}, {Cheek}, {De Angeli}, {Fabricius}, {Guerra}, {Holl}, {Masana},
  {Messineo}, {Mowlavi}, {Nienartowicz}, {Panuzzo}, {Portell}, {Riello},
  {Seabroke}, {Tanga}, {Th{\'e}venin}, {Gracia-Abril}, {Comoretto},
  {Garcia-Reinaldos}, {Teyssier}, {Altmann}, {Andrae}, {Audard},
  {Bellas-Velidis}, {Benson}, {Berthier}, {Blomme}, {Burgess}, {Busso},
  {Carry}, {Cellino}, {Clementini}, {Clotet}, {Creevey}, {Davidson}, {De
  Ridder}, {Delchambre}, {Dell'Oro}, {Ducourant},
  {Fern{\'a}ndez-Hern{\'a}ndez}, {Fouesneau}, {Fr{\'e}mat}, {Galluccio},
  {Garc{\'\i}a-Torres}, {Gonz{\'a}lez-N{\'u}{\~n}ez}, {Gonz{\'a}lez-Vidal},
  {Gosset}, {Guy}, {Halbwachs}, {Hambly}, {Harrison}, {Hern{\'a}ndez},
  {Hestroffer}, {Hodgkin}, {Hutton}, {Jasniewicz}, {Jean-Antoine-Piccolo},
  {Jordan}, {Korn}, {Krone-Martins}, {Lanzafame}, {Lebzelter}, {L{\"o}ffler},
  {Manteiga}, {Marrese}, {Mart{\'\i}n-Fleitas}, {Moitinho}, {Mora}, {Muinonen},
  {Osinde}, {Pancino}, {Pauwels}, {Petit}, {Recio-Blanco}, {Richards},
  {Rimoldini}, {Robin}, {Sarro}, {Siopis}, {Smith}, {Sozzetti}, {S{\"u}veges},
  {Torra}, {van Reeven}, {Abbas}, {Abreu Aramburu}, {Accart}, {Aerts},
  {Altavilla}, {{\'A}lvarez}, {Alvarez}, {Alves}, {Anderson}, {Andrei},
  {Anglada Varela}, {Antiche}, {Antoja}, {Arcay}, {Astraatmadja}, {Bach},
  {Baker}, {Balaguer-N{\'u}{\~n}ez}, {Balm}, {Barache}, {Barata}, {Barbato},
  {Barblan}, {Barklem}, {Barrado}, {Barros}, {Barstow}, {Bartholom{\'e}
  Mu{\~n}oz}, {Bassilana}, {Becciani}, {Bellazzini}, {Berihuete}, {Bertone},
  {Bianchi}, {Bienaym{\'e}}, {Blanco-Cuaresma}, {Boch}, {Boeche}, {Bombrun},
  {Borrachero}, {Bossini}, {Bouquillon}, {Bourda}, {Bragaglia}, {Bramante},
  {Breddels}, {Bressan}, {Brouillet}, {Br{\"u}semeister}, {Brugaletta},
  {Bucciarelli}, {Burlacu}, {Busonero}, {Butkevich}, {Buzzi}, {Caffau},
  {Cancelliere}, {Cannizzaro}, {Cantat-Gaudin}, {Carballo}, {Carlucci},
  {Carrasco}, {Casamiquela}, {Castellani}, {Castro-Ginard}, {Charlot},
  {Chemin}, {Chiavassa}, {Cocozza}, {Costigan}, {Cowell}, {Crifo}, {Crosta},
  {Crowley}, {Cuypers}, {Dafonte}, {Damerdji}, {Dapergolas}, {David}, {David},
  {de Laverny}, {De Luise}, {De March}, {de Martino}, {de Souza}, {de Torres},
  {Debosscher}, {del Pozo}, {Delbo}, {Delgado}, {Delgado}, {Di Matteo},
  {Diakite}, {Diener}, {Distefano}, {Dolding}, {Drazinos}, {Dur{\'a}n},
  {Edvardsson}, {Enke}, {Eriksson}, {Esquej}, {Eynard Bontemps}, {Fabre},
  {Fabrizio}, {Faigler}, {Falc{\~a}o}, {Farr{\`a}s Casas}, {Federici},
  {Fedorets}, {Fernique}, {Figueras}, {Filippi}, {Findeisen}, {Fonti},
  {Fraile}, {Fraser}, {Fr{\'e}zouls}, {Gai}, {Galleti}, {Garabato},
  {Garc{\'\i}a-Sedano}, {Garofalo}, {Garralda}, {Gavel}, {Gavras}, {Gerssen},
  {Geyer}, {Giacobbe}, {Gilmore}, {Girona}, {Giuffrida}, {Glass}, {Gomes},
  {Granvik}, {Gueguen}, {Guerrier}, {Guiraud}, {Guti{\'e}rrez-S{\'a}nchez},
  {Haigron}, {Hatzidimitriou}, {Hauser}, {Haywood}, {Heiter}, {Helmi}, {Heu},
  {Hilger}, {Hobbs}, {Hofmann}, {Holland}, {Huckle}, {Hypki}, {Icardi},
  {Jan{\ss}en}, {Jevardat de Fombelle}, {Jonker}, {Juh{\'a}sz}, {Julbe},
  {Karampelas}, {Kewley}, {Klar}, {Kochoska}, {Kohley}, {Kolenberg},
  {Kontizas}, {Kontizas}, {Koposov}, {Kordopatis}, {Kostrzewa-Rutkowska},
  {Koubsky}, {Lambert}, {Lanza}, {Lasne}, {Lavigne}, {Le Fustec}, {Le
  Poncin-Lafitte}, {Lebreton}, {Leccia}, {Leclerc}, {Lecoeur-Taibi},
  {Lenhardt}, {Leroux}, {Liao}, {Licata}, {Lindstr{\o}m}, {Lister}, {Livanou},
  {Lobel}, {L{\'o}pez}, {Managau}, {Mann}, {Mantelet}, {Marchal}, {Marchant},
  {Marconi}, {Marinoni}, {Marschalk{\'o}}, {Marshall}, {Martino}, {Marton},
  {Mary}, {Massari}, {Matijevi{\v{c}}}, {Mazeh}, {McMillan}, {Messina},
  {Michalik}, {Millar}, {Molina}, {Molinaro}, {Moln{\'a}r}, {Montegriffo},
  {Mor}, {Morbidelli}, {Morel}, {Morris}, {Mulone}, {Muraveva}, {Musella},
  {Nelemans}, {Nicastro}, {Noval}, {O'Mullane}, {Ord{\'e}novic},
  {Ord{\'o}{\~n}ez-Blanco}, {Osborne}, {Pagani}, {Pagano}, {Pailler},
  {Palacin}, {Palaversa}, {Panahi}, {Pawlak}, {Piersimoni}, {Pineau}, {Plachy},
  {Plum}, {Poggio}, {Poujoulet}, {Pr{\v{s}}a}, {Pulone}, {Racero}, {Ragaini},
  {Rambaux}, {Ramos-Lerate}, {Regibo}, {Reyl{\'e}}, {Riclet}, {Ripepi}, {Riva},
  {Rivard}, {Rixon}, {Roegiers}, {Roelens}, {Romero-G{\'o}mez}, {Rowell},
  {Royer}, {Ruiz-Dern}, {Sadowski}, {Sagrist{\`a} Sell{\'e}s}, {Sahlmann},
  {Salgado}, {Salguero}, {Sanna}, {Santana-Ros}, {Sarasso}, {Savietto},
  {Schultheis}, {Sciacca}, {Segol}, {Segovia}, {S{\'e}gransan}, {Shih},
  {Siltala}, {Silva}, {Smart}, {Smith}, {Solano}, {Solitro}, {Sordo}, {Soria
  Nieto}, {Souchay}, {Spagna}, {Spoto}, {Stampa}, {Steele},
  {Steidelm{\"u}ller}, {Stephenson}, {Stoev}, {Suess}, {Surdej}, {Szabados},
  {Szegedi-Elek}, {Tapiador}, {Taris}, {Tauran}, {Taylor}, {Teixeira},
  {Terrett}, {Teyssand ier}, {Thuillot}, {Titarenko}, {Torra Clotet}, {Turon},
  {Ulla}, {Utrilla}, {Uzzi}, {Vaillant}, {Valentini}, {Valette}, {van Elteren},
  {Van Hemelryck}, {van Leeuwen}, {Vaschetto}, {Vecchiato}, {Veljanoski},
  {Viala}, {Vicente}, {Vogt}, {von Essen}, {Voss}, {Votruba}, {Voutsinas},
  {Walmsley}, {Weiler}, {Wertz}, {Wevers}, {Wyrzykowski}, {Yoldas},
  {{\v{Z}}erjal}, {Ziaeepour}, {Zorec}, {Zschocke}, {Zucker}, {Zurbach}, \&
  {Zwitter}}]{gaia18}
{Gaia Collaboration}, {Brown}, A.~G.~A., {Vallenari}, A., {et~al.} 2018, \aap,
  616, A1, \dodoi{10.1051/0004-6361/201833051}

\bibitem[{{Gatley} {et~al.}(1987){Gatley}, {Hasegawa}, {Suzuki}, {Garden},
  {Brand}, {Lightfoot}, {Glencross}, {Okuda}, \& {Nagata}}]{gatley87b}
{Gatley}, I., {Hasegawa}, T., {Suzuki}, H., {et~al.} 1987, \apjl, 318, L73,
  \dodoi{10.1086/184940}

\bibitem[{{Goicoechea} {et~al.}(2016){Goicoechea}, {Pety}, {Cuadrado},
  {Cernicharo}, {Chapillon}, {Fuente}, {Gerin}, {Joblin}, {Marcelino}, \&
  {Pilleri}}]{goicoechea16}
{Goicoechea}, J.~R., {Pety}, J., {Cuadrado}, S., {et~al.} 2016, \nat, 537, 207,
  \dodoi{10.1038/nature18957}

\bibitem[{{Goicoechea} {et~al.}(2017){Goicoechea}, {Cuadrado}, {Pety}, {Bron},
  {Black}, {Cernicharo}, {Chapillon}, {Fuente}, \& {Gerin}}]{goicoechea17}
{Goicoechea}, J.~R., {Cuadrado}, S., {Pety}, J., {et~al.} 2017, \aap, 601, L9,
  \dodoi{10.1051/0004-6361/201730716}

\bibitem[{{Gould} \& {Salpeter}(1963)}]{gould63}
{Gould}, R.~J., \& {Salpeter}, E.~E. 1963, \apj, 138, 393,
  \dodoi{10.1086/147654}

\bibitem[{{Habart} {et~al.}(2011){Habart}, {Abergel}, {Boulanger}, {Joblin},
  {Verstraete}, {Compi{\`e}gne}, {Pineau Des For{\^e}ts}, \& {Le
  Bourlot}}]{habart11}
{Habart}, E., {Abergel}, A., {Boulanger}, F., {et~al.} 2011, \aap, 527, A122,
  \dodoi{10.1051/0004-6361/20077327}

\bibitem[{{Habart} {et~al.}(2005){Habart}, {Abergel}, {Walmsley}, {Teyssier},
  \& {Pety}}]{habart05}
{Habart}, E., {Abergel}, A., {Walmsley}, C.~M., {Teyssier}, D., \& {Pety}, J.
  2005, \aap, 437, 177, \dodoi{10.1051/0004-6361:20041546}

\bibitem[{{Habart} {et~al.}(2004){Habart}, {Boulanger}, {Verstraete},
  {Walmsley}, \& {Pineau des For{\^e}ts}}]{habart04}
{Habart}, E., {Boulanger}, F., {Verstraete}, L., {Walmsley}, C.~M., \& {Pineau
  des For{\^e}ts}, G. 2004, \aap, 414, 531, \dodoi{10.1051/0004-6361:20031659}

\bibitem[{{Habing}(1968)}]{habing1968}
{Habing}, H.~J. 1968, \bain, 19, 421
\newblock \url{https://ui.adsabs.harvard.edu/abs/1968BAN....19..421H}

\bibitem[{{Hasegawa} {et~al.}(1987){Hasegawa}, {Gatley}, {Garden}, {Brand},
  {Ohishi}, {Hayashi}, \& {Kaifu}}]{hasegawa87}
{Hasegawa}, T., {Gatley}, I., {Garden}, R.~P., {et~al.} 1987, \apjl, 318, L77,
  \dodoi{10.1086/184941}

\bibitem[{{Hayashi} {et~al.}(1985){Hayashi}, {Hasegawa}, {Gatley}, {Garden}, \&
  {Kaifu}}]{hayashi85}
{Hayashi}, M., {Hasegawa}, T., {Gatley}, I., {Garden}, R., \& {Kaifu}, N. 1985,
  \mnras, 215, 31P, \dodoi{10.1093/mnras/215.1.31P}

\bibitem[{{Hippelein} \& {M\"{u}nch}(1989)}]{hippelein89}
{Hippelein}, H.~H., \& {M\"{u}nch}, G. 1989, \aap, 213, 323
\newblock \url{https://ui.adsabs.harvard.edu/abs/1989A%26A...213..323H}

\bibitem[{{Hirota} {et~al.}(2008){Hirota}, {Ando}, {Bushimata}, {Choi},
  {Honma}, {Imai}, {Iwadate}, {Jike}, {Kameno}, {Kameya}, {Kamohara}, {Kan-Ya},
  {Kawaguchi}, {Kijima}, {Kim}, {Kobayashi}, {Kuji}, {Kurayama}, {Manabe},
  {Matsui}, {Matsumoto}, {Miyaji}, {Miyazaki}, {Nagayama}, {Nakagawa},
  {Namikawa}, {Nyu}, {Oh}, {Omodaka}, {Oyama}, {Sakai}, {Sasao}, {Sato},
  {Sato}, {Shibata}, {Tamura}, {Ueda}, \& {Yamashita}}]{hirota08}
{Hirota}, T., {Ando}, K., {Bushimata}, T., {et~al.} 2008, \pasj, 60, 961,
  \dodoi{10.1093/pasj/60.5.961}

\bibitem[{{Hollenbach} \& {Natta}(1995)}]{hollenbach95}
{Hollenbach}, D., \& {Natta}, A. 1995, \apj, 455, 133, \dodoi{10.1086/176562}

\bibitem[{{Hollenbach} \& {Salpeter}(1971)}]{hollenbach71}
{Hollenbach}, D., \& {Salpeter}, E.~E. 1971, \apj, 163, 155,
  \dodoi{10.1086/150754}

\bibitem[{{Hollenbach} {et~al.}(1991){Hollenbach}, {Takahashi}, \&
  {Tielens}}]{hollenbach91}
{Hollenbach}, D.~J., {Takahashi}, T., \& {Tielens}, A.~G.~G.~M. 1991, \apj,
  377, 192, \dodoi{10.1086/170347}

\bibitem[{{Hollenbach} \& {Tielens}(1999)}]{hollenbach99}
{Hollenbach}, D.~J., \& {Tielens}, A.~G.~G.~M. 1999, Reviews of Modern Physics,
  71, 173, \dodoi{10.1103/RevModPhys.71.173}

\bibitem[{{Howe} {et~al.}(1990){Howe}, {Jaffe}, \& {Geballe}}]{howe90}
{Howe}, J.~E., {Jaffe}, D.~T., \& {Geballe}, T.~R. 1990, in \baas, Vol.~22,
  Bulletin of the American Astronomical Society, 1328
  \newblock \url{https://ui.adsabs.harvard.edu/abs/1990BAAS...22.1328H}

\bibitem[{{Hurwitz}(1998)}]{hurwitz98}
{Hurwitz}, M. 1998, \apjl, 500, L67, \dodoi{10.1086/311387}

\bibitem[{{Joblin} {et~al.}(2018){Joblin}, {Bron}, {Pinto}, {Pilleri}, {Le
  Petit}, {Gerin}, {Le Bourlot}, {Fuente}, {Berne}, {Goicoechea}, {Habart},
  {K{\"o}hler}, {Teyssier}, {Nagy}, {Montillaud}, {Vastel}, {Cernicharo},
  {R{\"o}llig}, {Ossenkopf-Okada}, \& {Bergin}}]{joblin18}
{Joblin}, C., {Bron}, E., {Pinto}, C., {et~al.} 2018, \aap, 615, A129,
  \dodoi{10.1051/0004-6361/201832611}

\bibitem[{Kaplan {et~al.}(2017)Kaplan, Dinerstein, Oh, Mace, Kim, Sokal, Pavel,
  Lee, Pak, Park, Oh, \& Jaffe}]{kaplan17}
Kaplan, K.~F., Dinerstein, H.~L., Oh, H., {et~al.} 2017, The Astrophysical
  Journal, 838, 152
 \dodoi{10.3847/1538-4357/aa5b9f}
 

\bibitem[{{Le} {et~al.}(2016){Le}, {Pak}, {Kaplan}, {Mace}, {Lee}, {Pavel},
  {Jeong}, {Oh}, {Lee}, {Chun}, {Yuk}, {Pyo}, {Hwang}, {Kim}, {Park}, {Sok Oh},
  {Yu}, {Park}, {Minh}, \& {Jaffe}}]{le16}
{Le}, H.~A.~N., {Pak}, S., {Kaplan}, K.~F., {et~al.} 2016, \apj, 841, 13,
\dodoi{10.3847/1538-4357/aa6bf7}

\bibitem[{{Le Bourlot} {et~al.}(2012){Le Bourlot}, {Le Petit}, {Pinto},
  {Roueff}, \& {Roy}}]{lebourlot12}
{Le Bourlot}, J., {Le Petit}, F., {Pinto}, C., {Roueff}, E., \& {Roy}, F. 2012,
  \aap, 541, A76,
\dodoi{10.1051/0004-6361/201118126}

\bibitem[{{Le Petit} {et~al.}(2006){Le Petit}, {Nehm{\'e}}, {Le Bourlot}, \&
  {Roueff}}]{lepetit06}
{Le Petit}, F., {Nehm{\'e}}, C., {Le Bourlot}, J., \& {Roueff}, E. 2006, \apjs,
  164, 506, \dodoi{10.1086/503252}

\bibitem[{Lee(2015)}]{plp}
Lee, J.-J. 2015, plp: Version 2.0, \dodoi{10.5281/zenodo.18579}.

\bibitem[{{Luhman} {et~al.}(1997{\natexlab{a}}){Luhman}, {Jaffe}, {Sternberg},
  {Herrmann}, \& {Poglitsch}}]{luhman97b}
{Luhman}, M.~L., {Jaffe}, D.~T., {Sternberg}, A., {Herrmann}, F., \&
  {Poglitsch}, A. 1997{\natexlab{a}}, \apj, 482, 298, \dodoi{10.1086/304128}

\bibitem[{{Luhman} {et~al.}(1997{\natexlab{b}}){Luhman}, {Luhman}, {Benedict},
  {Jaffe}, \& {Fischer}}]{luhman97}
{Luhman}, M.~L., {Luhman}, K.~L., {Benedict}, T., {Jaffe}, D.~T., \& {Fischer},
  J. 1997{\natexlab{b}}, \apjl, 480, L133, \dodoi{10.1086/310627}

\bibitem[{{Marconi} {et~al.}(1998){Marconi}, {Testi}, {Natta}, \&
  {Walmsley}}]{marconi98}
{Marconi}, A., {Testi}, L., {Natta}, A., \& {Walmsley}, C.~M. 1998, \aap, 330,
  696
 \newblock \url{https://ui.adsabs.harvard.edu/abs/1998A%26A...330..696M}

\bibitem[{{Martini} {et~al.}(1999){Martini}, {Sellgren}, \&
  {DePoy}}]{martini99}
{Martini}, P., {Sellgren}, K., \& {DePoy}, D.~L. 1999, \apj, 526, 772,
  \dodoi{10.1086/308040}

\bibitem[{{Meixner} \& {Tielens}(1993)}]{meixner93}
{Meixner}, M., \& {Tielens}, A.~G.~G.~M. 1993, \apj, 405, 216,
  \dodoi{10.1086/172355}

\bibitem[{{Nishiyama} {et~al.}(2006){Nishiyama}, {Nagata}, {Kusakabe},
  {Matsunaga}, {Naoi}, {Kato}, {Nagashima}, {Sugitani}, {Tamura}, {Tanab{\'e}},
  \& {Sato}}]{nishiyama06}
{Nishiyama}, S., {Nagata}, T., {Kusakabe}, N., {et~al.} 2006, \apj, 638, 839,
  \dodoi{10.1086/499038}
  
  \bibitem[{{Nishiyama} {et~al.}(2009){Nishiyama}, {Tamura}, {Hatano}, {Kato},
  {Tanab{\'e}}, {Sugitani}, \& {Nagata}}]{nishiyama09}
{Nishiyama}, S., {Tamura}, M., {Hatano}, H., {et~al.} 2009, \apj, 696, 1407,
  \dodoi{10.1088/0004-637X/696/2/1407}

\bibitem[{{Panagia}(1973)}]{panagia73}
{Panagia}, N. 1973, \aj, 78, 929, \dodoi{10.1086/111498}

\bibitem[{{Park} {et~al.}(2014){Park}, {Jaffe}, {Yuk}, {Chun}, {Pak}, {Kim},
  {Pavel}, {Lee}, {Oh}, {Jeong}, {Sim}, {Lee}, {Nguyen Le}, {Strubhar},
  {Gully-Santiago}, {Oh}, {Cha}, {Moon}, {Park}, {Brooks}, {Ko}, {Han}, {Nah},
  {Hill}, {Lee}, {Barnes}, {Yu}, {Kaplan}, {Mace}, {Kim}, {Lee}, {Hwang}, \&
  {Park}}]{park14}
{Park}, C., {Jaffe}, D.~T., {Yuk}, I.-S., {et~al.} 2014, Proc. SPIE, 9147, 1,
  \dodoi{10.1117/12.2056431}

\bibitem[{{Parmar} {et~al.}(1991){Parmar}, {Lacy}, \& {Achtermann}}]{parmar91}
{Parmar}, P.~S., {Lacy}, J.~H., \& {Achtermann}, J.~M. 1991, \apjl, 372, L25,
  \dodoi{10.1086/186015}
  
 \bibitem[{{Pellegrini} {et~al.}(2007){Pellegrini}, {Baldwin}, {Brogan},
  {Hanson}, {Abel}, {Ferland}, {Nemala}, {Shaw}, \& {Troland}}]{pellegrini07}
{Pellegrini}, E.~W., {Baldwin}, J.~A., {Brogan}, C.~L., {et~al.} 2007, \apj,
  658, 1119, \dodoi{10.1086/511258}

\bibitem[{{Pellegrini} {et~al.}(2009){Pellegrini}, {Baldwin}, {Ferland},
  {Shaw}, \& {Heathcote}}]{pellegrini09}
{Pellegrini}, E.~W., {Baldwin}, J.~A., {Ferland}, G.~J., {Shaw}, G., \&
  {Heathcote}, S. 2009, \apj, 693, 285, \dodoi{10.1088/0004-637X/693/1/285}


\bibitem[{{Perryman} {et~al.}(1997){Perryman}, {Lindegren}, {Kovalevsky},
  {Hoeg}, {Bastian}, {Bernacca}, {Cr{\'e}z{\'e}}, {Donati}, {Grenon},
  {Grewing}, {van Leeuwen}, {van der Marel}, {Mignard}, {Murray}, {Le Poole},
  {Schrijver}, {Turon}, {Arenou}, {Froeschl{\'e}}, \& {Petersen}}]{perryman97}
{Perryman}, M.~A.~C., {Lindegren}, L., {Kovalevsky}, J., {et~al.} 1997, \aap,
  323, L49
  \newblock \url{https://ui.adsabs.harvard.edu/abs/1997A%26A...323L..49P}

\bibitem[{{Roueff} {et~al.}(2019){Roueff}, {Abgrall}, {Czachorowski},
  {Pachucki}, {Puchalski}, \& {Komasa}}]{roueff19}
{Roueff}, E., {Abgrall}, H., {Czachorowski}, P., {et~al.} 2019, \aap, 630, A58,
  \dodoi{10.1051/0004-6361/201936249}

\bibitem[{{Schlafly} {et~al.}(2014){Schlafly}, {Green}, {Finkbeiner}, {Rix},
  {Bell}, {Burgett}, {Chambers}, {Draper}, {Hodapp}, {Kaiser}, {Magnier},
  {Martin}, {Metcalfe}, {Price}, \& {Tonry}}]{schlafly14}
{Schlafly}, E.~F., {Green}, G., {Finkbeiner}, D.~P., {et~al.} 2014, \apj, 786,
  29, \dodoi{10.1088/0004-637X/786/1/29}

\bibitem[{{Sellgren}(1986)}]{sellegren86}
{Sellgren}, K. 1986, \apj, 305, 399, \dodoi{10.1086/164255}

\bibitem[{{Shaw} {et~al.}(2005){Shaw}, {Ferland}, {Abel}, {Stancil}, \& {van
  Hoof}}]{shaw05}
{Shaw}, G., {Ferland}, G.~J., {Abel}, N.~P., {Stancil}, P.~C., \& {van Hoof},
  P.~A.~M. 2005, \apj, 624, 794, \dodoi{10.1086/429215}

\bibitem[{{Shaw} {et~al.}(2009){Shaw}, {Ferland}, {Henney}, {Stancil}, {Abel},
  {Pellegrini}, {Baldwin}, \& {van Hoof}}]{shaw09}
{Shaw}, G., {Ferland}, G.~J., {Henney}, W.~J., {et~al.} 2009, \apj, 701, 677,
  \dodoi{10.1088/0004-637X/701/1/677}

\bibitem[{{Sheffer} {et~al.}(2011){Sheffer}, {Wolfire}, {Hollenbach},
  {Kaufman}, \& {Cordier}}]{sheffer11}
{Sheffer}, Y., {Wolfire}, M.~G., {Hollenbach}, D.~J., {Kaufman}, M.~J., \&
  {Cordier}, M. 2011, \apj, 741, 45, \dodoi{10.1088/0004-637X/741/1/45}

\bibitem[{{Shenavrin} {et~al.}(2011){Shenavrin}, {Taranova}, \&
  {Nadzhip}}]{shenavrin11}
{Shenavrin}, V.~I., {Taranova}, O.~G., \& {Nadzhip}, A.~E. 2011, Astronomy
  Reports, 55, 31, \dodoi{10.1134/S1063772911010070}

\bibitem[{{Sigut} \& {Jones}(2007)}]{sigut07}
{Sigut}, T.~A.~A., \& {Jones}, C.~E. 2007, \apj, 668, 481,
  \dodoi{10.1086/521209}

\bibitem[{{Sternberg}(1988)}]{sternberg88}
{Sternberg}, A. 1988, \apj, 332, 400, \dodoi{10.1086/166664}

\bibitem[{{Sternberg}(1989)}]{sterberg89b}
---. 1989, \apj, 347, 863, \dodoi{10.1086/168177}

\bibitem[{{Sternberg} \& {Dalgarno}(1989)}]{sternberg89}
{Sternberg}, A., \& {Dalgarno}, A. 1989, \apj, 338, 197, \dodoi{10.1086/167193}

\bibitem[{{Sternberg} {et~al.}(2014){Sternberg}, {Le Petit}, {Roueff}, \& {Le
  Bourlot}}]{sternberg14}
{Sternberg}, A., {Le Petit}, F., {Roueff}, E., \& {Le Bourlot}, J. 2014, \apj,
  790, 10, \dodoi{10.1088/0004-637X/790/1/10}

\bibitem[{{Sternberg} \& {Neufeld}(1999)}]{sternberg99}
{Sternberg}, A., \& {Neufeld}, D.~A. 1999, \apj, 516, 371,
  \dodoi{10.1086/307115}

\bibitem[{{St{\"o}rzer} \& {Hollenbach}(1998)}]{storzer98}
{St{\"o}rzer}, H., \& {Hollenbach}, D. 1998, \apj, 495, 853,
  \dodoi{10.1086/305315}

\bibitem[{{Takahashi} \& {Uehara}(2001)}]{takahashi2001}
{Takahashi}, J., \& {Uehara}, H. 2001, \apj, 561, 843, \dodoi{10.1086/323364}

\bibitem[{{Takami} {et~al.}(2000){Takami}, {Usuda}, {Sugai}, {Kawabata},
  {Suto}, \& {Tanaka}}]{takami00}
{Takami}, M., {Usuda}, T., {Sugai}, H., {et~al.} 2000, \apj, 529, 268,
  \dodoi{10.1086/308234}

\bibitem[{{Takayanagi} {et~al.}(1987){Takayanagi}, {Sakimoto}, \&
  {Onda}}]{takayanagi87}
{Takayanagi}, K., {Sakimoto}, K., \& {Onda}, K. 1987, \apjl, 318, L81,
  \dodoi{10.1086/184942}

\bibitem[{{Tanaka} {et~al.}(1989){Tanaka}, {Hasegawa}, {Hayashi}, {Brand}, \&
  {Gatley}}]{tanaka89}
{Tanaka}, M., {Hasegawa}, T., {Hayashi}, S.~S., {Brand}, P.~W.~J.~L., \&
  {Gatley}, I. 1989, \apj, 336, 207, \dodoi{10.1086/167006}

\bibitem[{{Thi} {et~al.}(2009){Thi}, {van Dishoeck}, {Bell}, {Viti}, \&
  {Black}}]{thi09}
{Thi}, W.-F., {van Dishoeck}, E.~F., {Bell}, T., {Viti}, S., \& {Black}, J.
  2009, \mnras, 400, 622, \dodoi{10.1111/j.1365-2966.2009.15501.x}

\bibitem[{{Tielens} \& {Hollenbach}(1985)}]{tielens85}
{Tielens}, A.~G.~G.~M., \& {Hollenbach}, D. 1985, \apj, 291, 722,
  \dodoi{10.1086/163111}

\bibitem[{{Timmermann} {et~al.}(1996){Timmermann}, {Bertoldi}, {Wright},
  {Drapatz}, {Draine}, {Haser}, \& {Sternberg}}]{timmermann96}
{Timmermann}, R., {Bertoldi}, F., {Wright}, C.~M., {et~al.} 1996, \aap, 315,
  L281
  \newblock \url{https://ui.adsabs.harvard.edu/abs/1996A%26A...315L.281T}

\bibitem[{{van Dokkum}(2001)}]{dokkum2001}
{van Dokkum}, P.~G. 2001, \pasp, 113, 1420, \dodoi{10.1086/323894}

\bibitem[{{Wang} \& {Chen}(2019)}]{wang19}
{Wang}, S., \& {Chen}, X. 2019, \apj, 877, 116,
  \dodoi{10.3847/1538-4357/ab1c61}

\bibitem[{{Warren} \& {Hesser}(1977)}]{warren77}
{Warren}, Jr., W.~H., \& {Hesser}, J.~E. 1977, \apjs, 34, 115,
  \dodoi{10.1086/190446}

\bibitem[{{Witt} {et~al.}(1989){Witt}, {Stecher}, {Boroson}, \&
  {Bohlin}}]{witt89}
{Witt}, A.~N., {Stecher}, T.~P., {Boroson}, T.~A., \& {Bohlin}, R.~C. 1989,
  \apjl, 336, L21, \dodoi{10.1086/185352}

\bibitem[{{Wu} {et~al.}(2018){Wu}, {Bron}, {Onaka}, {Le Petit}, {Galliano},
  {Languignon}, {Nakamura}, \& {Okada}}]{wu18}
{Wu}, R., {Bron}, E., {Onaka}, T., {et~al.} 2018, \aap, 618, A53,
  \dodoi{10.1051/0004-6361/201832595}

\end{thebibliography}

\end{document}